\documentclass[twoside,11pt]{article}

\usepackage[abbrvbib]{jmlr2e}
\usepackage{amssymb}
\usepackage{amsmath}
\usepackage{bm}
\usepackage{graphicx}
\usepackage[utf8x]{inputenc}
\usepackage{relsize}
\usepackage{bbm}
\usepackage{subcaption}
\usepackage{mathtools}
\usepackage{enumerate}
\usepackage{multirow}
\usepackage[table]{xcolor}
\usepackage{cleveref}
\usepackage{boldline}
\DeclareMathOperator*{\argmin}{argmin}

\renewcommand{\phi}{\varphi}

\newcommand{\R}{\mathbb{R}}

\jmlrheading{1}{2022}{1-48}{4/00}{10/00}{Chaumeny22a}{Chaumeny et al.}

\ShortHeadings{Bayesian nonparametric mixtures in practice}{Chaumeny et al.}
\firstpageno{1}

\begin{document}

\title{Bayesian nonparametric mixture inconsistency for the number  of components: How worried should we be in practice?}

\author{\name Yannis Chaumeny\thanks{Yannis Chaumeny and Johan van der Molen Moris contributed equally} \email yannis.chaumeny@alumni.epfl.ch \\
       \addr Department of Statistics\\
       Ecole Polytechnique Fédérale de Lausanne\\
       Lausanne, Switzerland
       \AND
       \name Johan van der Molen Moris$^\ast$ \email johan.vdmolen@mrc-bsu.cam.ac.uk \\
       \addr MRC Biostatistics Unit\\
       University of Cambridge\\
       Cambridge, UK
       \AND
       \name Anthony C. Davison \email anthony.davison@epfl.ch \\
       \addr Department of Statistics\\
       Ecole Polytechnique Fédérale de Lausanne\\
       Lausanne, Switzerland
       \AND
       \name Paul D. W. Kirk \email paul.kirk@mrc-bsu.cam.ac.uk \\
       \addr MRC Biostatistics Unit\\
       University of Cambridge\\
       Cambridge, UK
\\}

\editor{Kevin Murphy and Bernhard Sch{\"o}lkopf}

\maketitle

\begin{abstract}%   <- trailing '%' for backward compatibility of .sty file
We consider the Bayesian mixture of finite mixtures (MFMs) and Dirichlet process mixture (DPM) models for clustering. Recent asymptotic theory has established that DPMs overestimate the number of clusters for large samples and that estimators from both classes of models are inconsistent for the number of clusters under misspecification, but  the implications for finite sample analyses are unclear. The final reported estimate after fitting these models is often a single representative clustering obtained using an MCMC summarisation technique, but it is unknown how well such a summary estimates the number of clusters. Here we investigate these practical considerations through simulations and an application to gene expression data, and find that (i) DPMs overestimate the number of clusters even in finite samples, but only to a limited degree that may be correctable using appropriate summaries, and (ii) misspecification can lead to considerable overestimation of the number of clusters in both DPMs and MFMs, but results are nevertheless often still interpretable. We provide recommendations on MCMC summarisation and suggest that although the more appealing asymptotic properties of MFMs provide strong motivation to prefer them, results obtained using MFMs and DPMs are often very similar in practice.

\end{abstract}


\section{Introduction}
Many areas of applied statistics use mixture models for cluster analysis.
One advantage of this approach compared to heuristic counterparts is the possibility of inference on both individual cluster allocations and the entire clustering structure. Mixture models have been developed for a wide variety of data types, with applications to gene expression analysis \citep{medvedovic02,mclachlan02,desouto08,onogi11,crook18}, economics \citep{alfo08,wedel02,fruhwirth-schnatter04}, cell type classification \citep{chan08,prabhakaran16}, community detection \citep{geng19,legramanti21}, climate science \citep{gaffney07}, astrophysics \citep{mukherjee98}, health surveys \citep{ni20}, demographics \citep{carmona19} and  finance \citep{dias15}.
Reliable estimation of the number of clusters is essential for accurate prediction and decision-making---for example, precision medicine relies on identifying disease subtypes \citep{sorlie01,lanza13}---and various approaches to this have been developed \citep{dudoit02,james01,nobile04,woo06,henna08,celeux19}.%\footnote{Would it be useful to introduce some notation $K$ for the number of clusters/components already here, to shorten the text below?  Also the text slides from clusters to components, which may not be 100\% desirable?}

%Here we briefly mention what frequentists do
%In general, the number of components of a mixture model places an upper bound on the number of clusters, which correspond to components responsible for generating the data. It is usually assumed that the true number of clusters is equal to the true number of components, therefore, the terms ``estimation of number of clusters/components'' are used interchangeably, however, it is important to note the difference in the models. 
Frequentist estimation of the number of %components of a mixture model
clusters is commonly treated as a model selection problem, approaches to which include likelihood ratio tests \citep{mclachlan87}, information criteria \citep{biernacki00,celeux06,drton17} and overfitting with penalised likelihood mixture models \citep{manole21}. Identifiability issues are a major difficulty in both Bayesian and frequentist formulations \citep{liu03}; see \cite{celeux19} for an overview of the latter.

%Now we say why we think the Bayesian approach is better and introduce the two models we study
Bayesian mixture models treat the number of clusters as an unknown parameter for which principled uncertainty estimates may be obtained. We consider the mixture of finite mixtures \citep{nobile04,miller18} and Dirichlet process mixture models \citep{maceachern94,escobar95,neal00}, two common extensions of the finite mixture model that allow inference for the number of clusters. The first benefits from consistency results \citep{nobile94}, but estimation methods were inefficient until recently \citep{miller18}. The second is a nonparametric model that has become popular due to the existence of  efficient estimation algorithms \citep{neal00,jain04}.

%Here we mention the negative theorical results
Despite their attractions, recent theory has established that estimators of the numbers of clusters based on these models have undesirable asymptotic properties. Dirichlet process mixtures have been shown to give inconsistent estimators of the number of clusters, overestimation of which has been observed for finite samples due to the appearance of small extra clusters \citep{miller14}. Mixtures of finite mixtures have been suggested as a consistent alternative, but model misspecification can still result in inconsistency and poor estimation of the number of clusters in practice \citep{cai21}. Moreover, high-dimensional data are particularly challenging for all clustering methods, which often incorrectly estimate the number of clusters \citep{chandra21}. 

%We are interested in what happens in finite samples
Although unsettling, poor asymptotic properties need not entail major problems in practical situations with finite samples, and our first aim in this work is to investigate to what extent the  asymptotic results sabotage the estimation of the number of clusters with Bayesian mixtures models with sample sizes that are typical for applications.

%Also, we use MCMC and summarisation methods, and we investigate their impact on final results
In practice inference is generally performed using Markov chain Monte Carlo (MCMC) algorithms \citep{neal00}. This results in a large number of MCMC samples of cluster allocations, which cannot be interpreted directly owing to the presence of multiple modes over all the partitions of the data, changes in the number of components, and label switching, so  summarisation methods have been introduced to find a representative clustering \citep{binder78,medvedovic04,fritsch09,wade18,rastelli18,dahl21}. To summarise MCMC samples despite label switching \citep{jasra05}, these post-processing methods are often based on Bayesian decision theory, that is, they minimise the posterior expectation of a suitable loss function.
The different choices among all possible combinations of loss function and optimisation method have led to the use of various summaries \citep{huelsenbeck07,savage10,kirk12,lock13,roth14,yuan15,carmona19,zafar19,fuentes-garcia19,ni20}, but the effect of summarisation methods has not yet been studied, especially with regard to the summarising number of clusters, which need not be representative of the corresponding marginal posterior distribution.
Hence a second goal of our work is to investigate how the summarisation method affects the number of clusters in light of the theoretical results. We focus on the low-dimensional case to avoid complications that can arise with clustering methods and MCMC samplers in high dimensions.

%Alternative methods and why we didn't study them
Another popular approach to estimating the number of components  is \emph{overfitting}\ with finite mixture models. By deliberately fitting a model with an upper bound $K$ on the number of components we can estimate the true number of clusters $K_0$ by discarding empty components.
\cite{rousseau11} show that overfitting can give  consistent estimators %of the number of clusters
, and that additional components vanish a posteriori at rate $N^{-1/2} \log(N)^\beta$ for some $\beta>0$. %depending on the Dirichlet prior hyperparameter $\gamma$, the dimension $p$ and $K-K_0$, the difference between the fitted and true number of components.
We do not consider this approach, which requires %an additional 
setting parameter $K$ which should be higher than the unknown $K_0$, and for $K$ high enough, this model is comparable to a Dirichlet process mixture \citep{fruhwirth-schnatter19}.
%However, under misspecification of the component distribution, \cite{cai21} show that the posterior number of clusters for finite mixture models converges to the maximum $K$. We did not implement this approach, so it remains unclear how summary clusterings affect the number of clusters.

Non-decision-theoretic approaches intended to correct Dirichlet process mixture overestimation of the number of components also exist. For instance, \cite{guha19} introduce an algorithm that merges small additional clusters of each MCMC sample in post-processing. They prove that their method is consistent for the number of components, but it involves a model-specific theoretical contraction rate to adjust the truncation of small clusters; they provide this rate for Gaussian mixture models. We do not consider such algorithms, because decision theory-based summaries are more common in practice.

%Description of the rest of the paper
 In Section~\ref{background}, we introduce the types of Bayesian mixture models that we consider, summarise known asymptotic (in)consistency results and present different strategies for summarising simulation output for these models. We perform a study on synthetic data in Section~\ref{chap:results} and illustrate the results using gene expression data. We conclude by discussing our results and give an overview of alternative approaches and potential future work.

\section{Background, theory and summarisation methods}\label{background}

\subsection{Bayesian mixture models}
\label{chap:mixture}

\subsubsection{Finite mixture model}
\label{sec:fmm}

Mixture models presuppose that the data are generated from a mixture of component distributions belonging to a given parametric family, and clusters are determined by observations generated by the same component. %Now we introduce the formal definition of the mixture models we study.
 A Bayesian finite mixture model for observations $X_1, \ldots,  X_N$ with $K$ components can be defined hierarchically as
\begin{equation}
    \begin{aligned}
    \label{def:fmm}
    X_i \mid \bm \theta, z_i &\sim F( \theta_{z_i} ), \quad i = 1, \ldots, N,\\
     \theta_k  &\sim H, \quad k = 1,\ldots,K, \\
         z_i \mid  \bm{\pi} &\sim \text{Categorical}_K(\bm{\pi}), \quad i = 1, \ldots, N, \\
    \bm{\pi}  &\sim \text{Dirichlet}_K(\gamma, \ldots , \gamma),
   \end{aligned}
\end{equation}
where the {\em component distribution} $F$ is parameterised by $\theta \in \Theta$ and $H$ is a prior distribution on the component parameters%, sometimes called the \emph{base measure}
. The {\em component allocations} $z_i$ indicate the component responsible for each observation, while the vector $\bm \pi$ represents the component proportions (also called the \emph{mixing distribution}), i.e., $\pi_k$ is interpreted as the prior probability that an observation belongs to component $k$. The prior on $\bm \pi$ is here taken to be a symmetric Dirichlet distribution with a parameter $\gamma > 0$ commonly set to $1$ or $1/K$. 
By marginalising over $\bm z$, the probability density function of an observation $X_i$ can be written as
\begin{equation}
\label{eq:mix}
p \left( X_i \mid \bm \theta, \bm \pi \right) = \sum_{k=1}^K \pi_k f( X_i \mid \theta_k),
\end{equation}
where $f$ is the component probability density function associated with $F$. Thus, we recover the standard representation of a mixture.% with a mixing distribution $\bm \pi$ satisfying $0 \leq \pi_k \leq 1$ for all $k=1, \ldots, K$ and $\sum_{k=1}^K \pi_k = 1$.

The most common choice of component distribution for continuous data  is multivariate Gaussian. For variables $ X_i \in \mathbbm R^p$ this corresponds to setting $F(\theta_k) = \mathcal N ( \mu_k,  \Sigma_k)$ such that $\theta_k = ( \mu_k, \Sigma_k)$ for $k=1, \ldots, K$, where $ \mu_k$ is a $p \times 1$ vector and $ \Sigma_k$ is a positive semi-definite $p \times p$ matrix.

\subsubsection{Mixture of finite mixtures model}
\label{sec:mfm}

A natural extension of the finite mixture model~(\ref{def:fmm}) is to treat the number of components $K$ as an unknown parameter with a prior $p_K$. The resulting mixture of finite mixtures (MFM) model can be defined hierarchically as
\begin{equation}
    \begin{aligned}
    \label{def:mfm}
    K &\sim p_K, \\
    \bm{\pi} \mid K &\sim \text{Dirichlet}_K(\gamma,\ldots,\gamma), \\
    z_i \mid K, \bm{\pi} &\sim \text{Categorical}_K(\bm{\pi}), \quad i = 1, \ldots, N, \\
    \theta_k \mid K &\sim H, \quad  k = 1,\ldots,K, \\
    X_i \mid K, \bm{\theta}, z_i &\sim F( \theta_{z_i} ), \quad  i = 1, \ldots, N.
    \end{aligned}
\end{equation}
If we set $p_K(k) = \mathbbm 1_{ \{ k = K_0 \} }$ we recover a finite mixture model with $K_0$ components. The mixture of finite mixtures model is sometimes also called a finite mixture model with a prior on the number of components \citep{nobile94,richardson97}. %The name mixture of finite mixtures comes from writing the likelihood as
% \begin{equation}
%     p( \bm X \mid \bm \theta, \bm \pi, \bm z ) = \sum_{k=1}^\infty p_K(k) p( \bm X \mid \bm \theta, \bm \pi, \bm z, K = k),
% \end{equation}
% such that we obtain the likelihood $p( \bm X \mid \bm \theta, \bm z, K = k)$ of a finite mixture with $k$ components. 
  Some examples of discrete priors $p_K$ are a Poisson distribution for $K-1$ \citep{nobile05} or a Geometric distribution \citep{miller18}.

% Since the number of components can change, the parameters $\bm \theta$ and $\bm z$ have an ambiguous definition. For example, a component parameter $\theta_2$ has a different dimension depending on whether $K=2$ or $K=3$, as noticed by \cite{nobile04}. In fact, given any value of $K$ an implicit collection of $K$ component parameters $\bm \theta$ and $N$ allocation parameters $\bm z$ is introduced, making the rigorous parameterisation of the mixture of finite mixtures model infinite-dimensional. %Moreover, the mixture of finite mixtures model inherits non-identifiability from the finite mixture model.

% [gaussian mixture sieve for frequentists?]

\subsubsection{Dirichlet process mixture model}

Another extension of the finite mixture model is the Dirichlet process mixture model, under which 
% This is one of the most popular nonparametric models, as it naturally captures a clustering distribution in a population.
each observation $X_i$ has an associated parameter, $\theta_i$, with $\theta_1, \ldots, \theta_N$ assumed to be independent and identically distributed according to a random distribution $G$.
The prior on $G$ is a Dirichlet process, $G \sim \text{DP} (\alpha, H)$, where $\alpha$ is the \emph{concentration parameter}\ and $H$ is the \emph{base measure}.\  %On marginalising the random distribution $G$, the $\theta_i$ become dependent and distributed as
The Dirichlet process mixture (DPM) model can therefore be written hierarchically as
\begin{equation}
    \begin{aligned}
    \label{def:dpm}
   G &\sim \text{DP}(\alpha, H),\\
    \theta _i \mid G &\sim G, \quad i = 1,\ldots,N, \\
     X_i \mid \theta_i &\sim F( \theta_i ), \quad  i = 1, \ldots, N.
    \end{aligned}
\end{equation}
A practical definition of the resulting distribution of $\theta$ is provided by the Pólya urn scheme \citep{blackwell73}.

In contrast to the MFM, under which the prior density of the number of components is fixed as $p_K$, the Dirichlet process prior allows an increasing number of components as the number of observations $N$ grows. Indeed, the prior expectation of $T$, the number of components for $X_1, \ldots, X_N$ or equivalently the number of unique values of $\bm z$, satisfies
\begin{equation}
    \mathbbm E (T) = \sum_{i = 1}^N \frac{\alpha}{\alpha + i -1} = O\left(\alpha \log \frac{N}{\alpha} \right).
\end{equation}
For more properties of  the Dirichlet process see \cite{ghosal10}, for example.
%  An underlying assumption of the Dirichlet process mixture model is therefore that an increasing population cannot be asymptotically captured by a finite number of components. In other words, as $N \rightarrow  \infty $ the Dirichlet process prior puts zero mass on finite mixtures. The prior growth rate of the number of components is determined by $\alpha$, and choosing $\alpha$ smaller can inhibit the creation of new clusters.

\subsection{Known asymptotic properties}

There are various theoretical results for the estimation of the number of components with Bayesian mixture models. In general we present asymptotic results as the number of observations $N$ tends to infinity. %, except in Section~\ref{subsec:chandra}, where we consider high-dimensional asymptotics.
Due to the generality of mixture models, useful results require further assumptions, but these hold in most practical settings and in particular for those we study in Section~\ref{chap:results}.

Although we focus on component allocation and the number of components, an underlying issue is density estimation: although the posterior number of components may be inconsistent, the mixture density $\sum_{k=1} \pi_k f(\bm X \mid  \theta_k)$ can still converge to the true generating density.   \cite{nguyen13}, for example, shows that the posterior distribution of a DPM model converges to the true distribution in the Wasserstein metric, and \cite{guha19} show that the posterior distribution of a MFM model contracts to the true generating parameters at an optimal rate under correct specification, up to relabeling of the components.  These authors also show that even under misspecification both types of models can asymptotically recover the true parameters under certain conditions.
Under an identifiability condition on the true generating distribution, they show in particular that a multivariate normal mixture distribution converges in the Wasserstein metric, though the convergence is slower than for the well-specified case. However, as we describe in Section~\ref{subsec:miss}, under misspecification the estimated number of components is  inconsistent and diverges. This is a recurring trade-off for mixture models: it is possible to obtain an arbitrarily good density estimate, but at the cost of overestimating the number of components.

\subsubsection{Consistency of mixture of finite mixtures}
\label{subsec:mfm_cons}

For the mixture of finite mixtures model, \cite{nobile94} shows that the posterior converges to the true number of components, if the mixture is identifiable and the component distribution is continuous. Mixture-identifiability means that a mixture of $K$ components is strictly more expressive than any mixture with $K' < K$ components and it is uniquely identified up to relabeling of the components. Many models satisfy this condition, including Gaussian mixtures and most common continuous component distributions.
%To present his result we need an identifiability assumption on the true generating distribution.

%\begin{definition}
%\label{def:mix_identifiable}
%    A finite mixture model with component distribution $F$ is called \textbf{mixture-identifiable} if $\sum_{j=1}^K\pi_j F\left( \theta_j\right)  = \sum_{j=1}^{K'} \pi'_{j}  F( \theta'_{j} )$ implies $K = K'$ and $\pi_j = \pi'_{\sigma(j)}, \theta_j =  \theta'_{\sigma(j)}$ for all $j = 1, \ldots , K$ and for some permutation $\sigma$.
%\end{definition}

%\begin{theorem} \citep{nobile94}\\
%\label{thm:mfm}
%Let $ X_1, \ldots, X_N$ follow a mixture-identifiable finite mixture model with $K_0$ components such that the component distribution $F$ is continuous.
%Suppose we model $X_1, \ldots, X_N$ by a mixture of finite mixtures with a prior $p_K$ on the number of components such that $p_K(K_0) > 0$. Then the posterior number of components converges to the true number of components, i.e.,
%\begin{equation}
%    \lim_{N \rightarrow \infty} p( K = K_0 \mid  X_1, \ldots , X_N) = 1 \quad \text{almost surely.}
%\end{equation}
%\end{theorem}
%The proof of Proposition~\ref{thm:mfm} is based on a general theorem by \cite{doob48} giving the posterior consistency of a large number of identifiable models. We refer to \cite{miller18d} for a more general statement of Doob's theorem and a complete proof. As another consequence of Doob's theorem, \cite{nobile94} shows that the component parameters $\bm \theta$ and the proportions $\bm \pi$ also have consistent posterior distributions up to relabeling.

\subsubsection{Inconsistency of Dirichlet process mixture}
\label{subsec:dpm_incons}

%In the previous section we summarised consistency results for the number of components for mixtures of finite mixtures.
%Naturally it is desirable to obtain a similar result for the nonparametric Dirichlet process prior.
%However, 
\cite{miller14} show that the posterior distribution of the number of components of a Dirichlet process mixture, or more generally a Pitman--Yor mixture model does not contract to the true number of components when the observations arise from a finite mixture. 
%\begin{theorem} \label{thm:mh} \citep{miller14} \\
%Let $X_1, \ldots, X_N$ be observations of a finite mixture whose component distribution $F$ is among the following:
%\begin{itemize}
%    \item multivariate normal $\mathcal N_p (\mu, \Sigma)$,
%    \item univariate $\text{Exponential}(\lambda), \text{Gamma}(\alpha,\beta), \text{Log-Normal}\left(\mu, \sigma^2 \right) \hspace{-1mm}\text{ or Weibull(a,b)}$ with fixed shape $a$.
%\end{itemize}
%Suppose we model $\bm X$ with a Dirichlet process mixture model with parameter $\alpha > 0$ with
%\begin{itemize}
%    \item a conjugate prior on the component distributions,
%    \item a discrete prior with non-empty support, or
%    \item any continuous prior density bounded and non-zero on some compact set.
%\end{itemize}
%Then, for any $t \in \mathbbm N_{\geq 1}$, 
%
%\begin{equation}
%\label{eq:mh}
%    \limsup_{n \rightarrow \infty} p( T_n = t \mid  X_1, \ldots,  X_N) < 1.
%\end{equation}
%
%\end{theorem}
%\cite{miller14} give a more general result that applies both to Pitman--Yor process priors --- a generalisation of the Dirichlet process --- and to a larger class of component distributions under more detailed assumptions.
%We give Proposition~\ref{thm:mh} with stronger assumptions which apply to various popular distributions.
%
%From the asymptotic result (\ref{eq:mh}) we deduce that the posterior distribution of the number of components does not contract to the true number of components.
 However it is still unclear what the posterior distribution actually does, for instance, it might contract around a wrong value, or diverge entirely. \cite{yang19} show that the posterior diverges when a uniform or Gaussian prior is taken for $\bm \theta$.
Since we do not know the convergence speed it is difficult to assess the finite-sample repercussions of this, and other aspects such as the choice of $\alpha$ may be more important when $N$ is finite.
In fact, \cite{miller14} do not cover the case where $\alpha$ is also a parameter with a hyperprior, though they suspect their result to hold regardless. However, more recent work has shown that consistency can be achieved by putting a prior on $\alpha$ \citep{ascolani_clustering_2022}, or allowing it to depend on the sample size \citep{ohn_optimal_2022}, although it is unclear whether these results hold for mixtures of multivariate Gaussian densities with unknown covariance matrix, which are most commonly used for clustering.
 
\subsubsection{Inconsistency under misspecification}
\label{subsec:miss}

The consistency result for the mixture of finite mixtures model in Section~\ref{subsec:mfm_cons} holds when the generating and modelling component distribution families are the same, but we may ask whether it remains true under misspecification, which is frequent, perhaps even invariable, in practice.
We saw in Section~\ref{subsec:dpm_incons} that misspecification of the allocation prior can lead to inconsistency. \cite{cai21} show that mixtures of finite mixtures with misspecified component distributions are also inconsistent for the number of components. Indeed, they show that the posterior diverges for any misspecified location-scale family of component distributions, if the corresponding mixture is identifiable and absolutely continuous with respect to the parameter $ \bm \theta$, and the data generating mixture lies in the Kullback--Liebler support of the prior. Under an assumption of degenerate limits they also prove a more general statement for any continuous density/mass function.
\label{chap:clust}

\subsection{Summarisation of MCMC samples}
\label{sec:mcmc}

Analytic computations for Bayesian mixture models are intractable, so the usual approach is to use Markov chain Monte Carlo (MCMC) sampling, which results in a time series $\bm z^{(1)},\ldots,\bm z^{(M)}$ from the posterior $p(\bm z,\bm  \theta \mid X_1, \ldots, X_N)$, where $M$ is chosen adequately large in practice.
Even if the $\bm z^{(m)}$ are sampled from the exact posterior distribution, the results cannot be interpreted directly  using sample averages or other standard summary statistics owing to two major  identifiability issues for the components, namely label-switching and variation in the number of components. %as the samples consist of an $N \times M$ matrix, which is usually large, so we seek a practical summary of the MCMC samples for interpretation and decision-making.
%A standard approach is to simply average the samples as a point estimate of the posterior mean. For component allocation samples this is not possible

Label-switching creates an incoherence in the values of the posterior samples of $\bm z$. Any permutation of the component labels leaves the posterior density unchanged,
so taking a summary clustering, such as the average of the samples, is not helpful, because  two samples $\bm z^{(m)}$ and $\bm z^{(m')}$ may represent the same component allocation with permuted labels: with $N=3$ and $K=2$, for example, the labelings $1,1,2$ and $2,2,1$ lead to the same clustering. %Figure~\ref{fig:peanuts} depicts label-switching during MCMC sampling for illustrative bivariate observations.
% \begin{figure}
%     \centering
%     \fbox{\includegraphics[width=0.7\linewidth]{figures/peanuts.png}}
%     \caption{Illustration of label-switching and varying number of clusters for three illustrative MCMC samples of $\bm z$. Colours represent the cluster allocation $\bm z$.}
%     \label{fig:peanuts}
% \end{figure}
Moreover, the number of components vary during sampling, further complicating the challenge of summarisation. %changing the meaning of a particular label, so %Indeed, assume for example that an observation $i$ has a label $z_i^{(m)} = 2$ for some sample $m$ with three components. Then if $z_i^{(n)} = 2$ for a different sample $n$ with five components, the same value $2$ refers to entirely different partitions. For example, in Figure~\ref{fig:peanuts} the red component label changes meaning from the first to second iteration to illustrate this phenomenon --- and in fact any component label is valid for the second sample.
%even if we could solve label-switching issues, we could not directly merge samples with different numbers of clusters into a summary based on the labels. We should therefore consider the component allocation samples only as partitions of the set $\{1, \ldots, N\}$, which is the only invariant of interest across samples.\footnote{?}

\begin{figure}
    \centering
    \fbox{\includegraphics[width=0.6\linewidth]{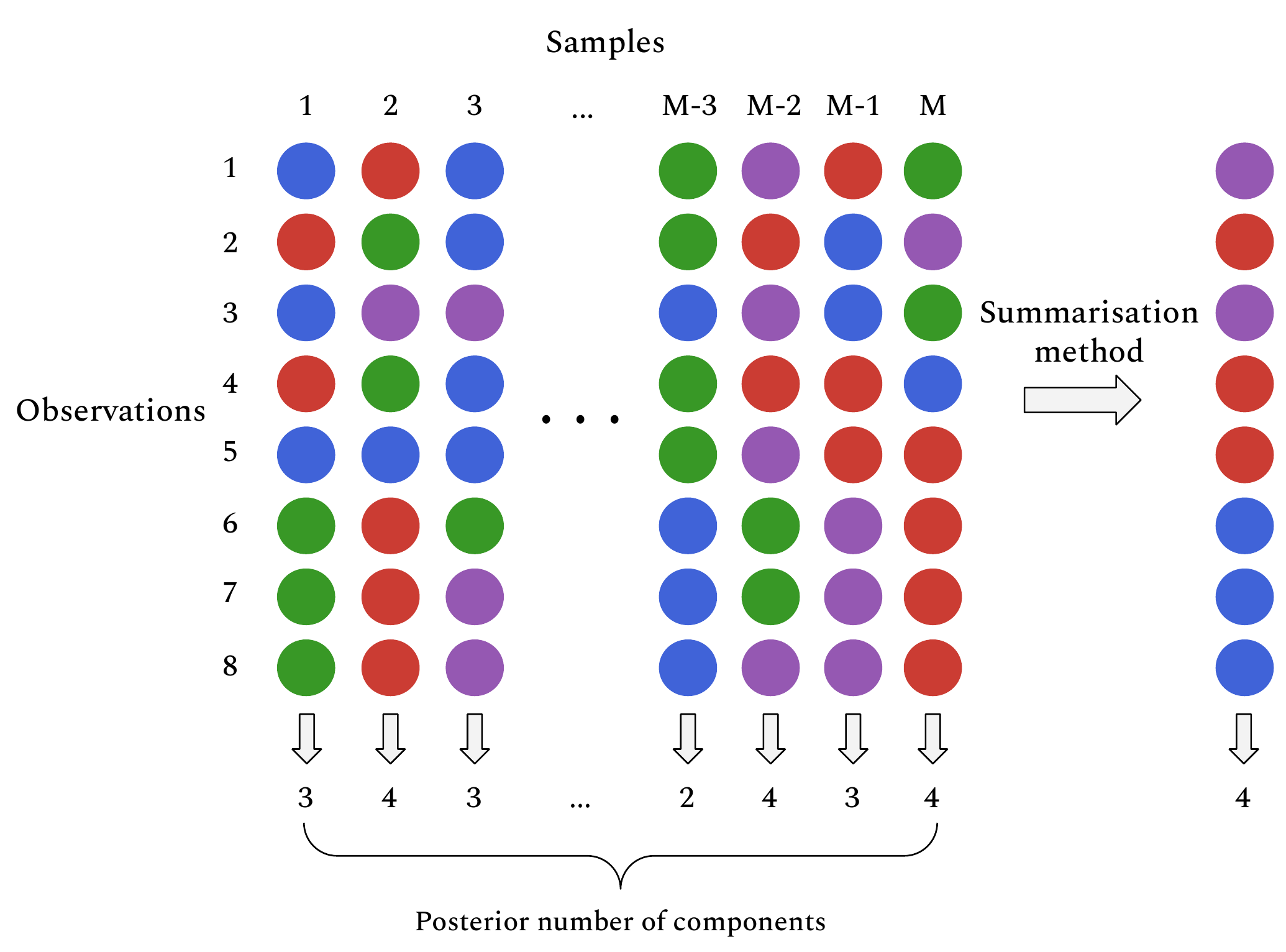}}
    \caption{Illustration of MCMC samples and an MCMC summary clustering. Cluster labels are represented by different colours.}%Needs more description
    \label{fig:summ_clust}
\end{figure}

Formally, a summarisation method is a function from $\bm z^{(1)},\ldots,\bm z^{(M)}$ to a single summary clustering $\bm z^*$ of the $N$ observations, i.e.,
\begin{equation}
\label{def:summ_clust}
\begin{aligned}
    \mathbbm N^{N\times M} &\rightarrow \mathbbm N ^N\\
    \big\{\bm  z^{(m)}  \big\}_{m = 1, \ldots, M } &\rightarrow \bm z^*.
\end{aligned}
\end{equation}
Figure~\ref{fig:summ_clust} illustrates the summarisation task and shows the difference between the posterior number of components and the number of clusters of a summary clustering.  Post-processing methods are intended to provide a ``representative'' $\bm z^*$.  The concept of representativity is not uniquely defined, but \cite{binder78} introduced a formal definition within a loss minimisation framework. For most summarisation methods we can define a loss function $L : \mathbbm N^N \times  \mathbbm N^N \rightarrow \R_{\geq 0}$ such that 
\begin{equation}
\label{def:loss}
    \bm z^* = \argmin_{\hat{\bm z}} \mathbbm E \left[ L(\bm z,\hat{\bm z}) \mid \bm X \right],
\end{equation}
where the expectation is taken over $\bm z$.
In order that the loss $L$ be invariant to relabeling, clusterings should be treated as partitions of the set $\{1, \ldots,N \}$, and then a direct MCMC approximation of~\eqref{def:loss} is
\begin{equation}
\label{eq:loss_approx}
    \bm z^* = \argmin_{\hat{\bm z}} \frac{1}{M} \sum_{m = 1}^M L \left( \bm z^{(m)},\hat{\bm z} \right). %\tilde p \left[ \bm z^{(m)} \mid \bm X \right],
\end{equation}
A central quantity for post-processing samples is the \emph{posterior similarity matrix}\ (PSM) $\bm P$ with elements
\begin{equation}
    \label{def:psm}
    P_{ij} = p( z_i = z_j \mid X ) 
    \approx \frac{1}{M} \sum_{m=1}^M \mathbbm 1 \left[ z_i^{(m)} = z_j^{(m)} \right], \quad i,j = 1, \ldots , N,
\end{equation}
where $\mathbbm 1$ is the indicator function, which measures the frequency of co-clustering of pairs of observations and is invariant to label-switching or to changes in the numbers of components during sampling.
To simplify notation in this section we write $K(\bm z)$ for the number of clusters in $\bm z$ and denote the contingency counts of two clusterings $\bm z$ and $\hat{\bm  z}\bm $ as
\begin{equation}
    n_{kk'} = \sum_{i,j = 1}^N \mathbbm 1(z_i = k) \mathbbm 1(\hat z_j = k'), \quad
    n_k = \sum_{i=1}^N \mathbbm 1(z_i = k),\quad k = 1, \ldots, K(\bm z), k ' = 1, \ldots, K(\hat{\bm z}).
\end{equation}
These form a $K(\bm z) \times K (\hat{\bm z})$ matrix representation of the information relevant to cluster comparisons.

In the following section we present various loss functions including the most used in practice. Given a loss function, we then need to optimise (\ref{def:loss}) to obtain a summary clustering. We present algorithms for approximate optimisation in Section~\ref{sec:opt}.

\subsection{Loss functions}

\subsubsection{Maximum a posteriori}

A straightforward approach to obtaining a summary clustering $\bm z^*$ is to take the sample $\bm z^{(m)}$ that maximizes the unnormalised posterior density. This corresponds to taking the 0-1 loss in (\ref{eq:loss_approx}), i.e., $L_{\text{0-1}}(\bm z,\hat{ \bm z}) = \mathbbm 1 ( \hat{\bm  z} = \bm z)$ where $\hat{\bm z} = \bm z$ is defined up to relabeling. This entirely circumvents the label switching problem by only considering a single sample, rather than comparing clusterings with possibly different labelings.

\subsubsection{Binder's loss}

The first loss function for summarising MCMC samples is due to \cite{binder78}, which uses pairwise co-clustering disagreement to measure differences between two clusterings via
\begin{equation}
    L_{\text{B}} (\bm z, \hat{\bm z}) = \mathlarger{\sum}_{i < j} 
    l_1 \mathbbm 1 (z_i = z_j) \mathbbm 1 ( \hat z_i \neq \hat z_j)
    + l_2 \mathbbm 1 (z_i \neq z_j) \mathbbm 1 (\hat z_i = \hat z_j),
\end{equation}
where $l_1, l_2 > 0 $ are commonly set to $l_1=l_2=1$, though other values could be used in case of asymmetric classification error. We set $l_1 = l_2 = 1$, but \cite{dahl21} show that changing the ratio $l_1/l_2$ can significantly affect the number of clusters.

The expected posterior loss can be written in terms of the posterior similarity matrix as
\begin{equation}
    \mathbbm E \left[ L_{\text{B}} (\bm z, \hat{\bm z} ) \mid X \right] = \sum_{i < j} \lvert \mathbbm 1 ( \hat z_i = \hat z_j) - P_{ij} \rvert .
\end{equation}
%  We can then use the approximation of the posterior similarity matrix from the samples as in (\ref{def:psm}).
Binder's loss has a direct relationship with the commonly-used Rand index $\text R(\cdot,\cdot)$  \citep{rand71}. Indeed, for $l_1 = l_2 = 1$,
\begin{equation}
\label{eq:binder_rand}
   L_{\text{B}} (\bm z, \hat{\bm z}) = {N \choose 2} \left[ 1 - \text R(\bm z, \hat{\bm z}) \right], 
\end{equation}
so minimising the expected posterior Binder loss is equivalent to maximising the expected posterior Rand index.

\subsubsection{Posterior Expected Adjusted Rand (PEAR)}

 \cite{fritsch09} propose using the  summary clustering that maximises the posterior adjusted Rand index \citep{hubert85}. While they do not explicitly introduce a loss function, we define the adjusted Rand loss as
\begin{equation}
    L_{\text{AR}} (\bm z, \hat{\bm z}) = 1 - \text{AR}(\bm z, \hat{\bm z}),
\end{equation}
where the adjusted Rand index AR$(\cdot,\cdot)$  corrects the Rand index for chance under permutations within clusters.
For computational reasons \cite{fritsch09} suggest taking the approximation $\mathbbm 1 (z_i = z_j) \approx P_{ij}$ in AR$(\bm z, \hat{\bm z})$, but \cite{dahl21} argue that this  is unnecessary and may lead to inaccuracies. Nevertheless, we will use %refer to\footnote{Do you mean use the approximate optimisation?}
the approximate optimisation, as it is typically used in practice.

\subsubsection{Variation of Information}
A loss function can also be based on a distance metric on the space of clusterings. \cite{wade18} suggest using as a loss function the variation of information \citep{meila07}, which has the advantage of inheriting theoretical properties from the distance space. The variation of information distance between two clusterings is
\begin{equation}
\hspace{-3mm}
\begin{aligned}
    L_\text{VI}(\bm z,\hat{\bm z}) &= -H(\bm z) - H( \hat{\bm  z}) + 2 H(\bm z, \hat{\bm z}) \\
    &=  \sum_{k=1}^{K(\bm z)} \frac{n_k}{N} \log_2\left(\frac{n_k}{N}\right) +
        \sum_{k'=1}^{K(\hat{\bm z}) } \frac{\hat n_{k'}}{N} \log_2\left(\frac{\hat n_{k'}}{N}\right)
       - 2 \sum_{k=1}^{K(\bm z)} \sum_{k'=1}^{K (\hat{\bm z})} \frac{n_{kk'}}{N} 
           \log_2\left(\frac{n_{kk'}}{N}\right),
\end{aligned}
\end{equation}
%where $\log_2$ is the base $2$ logarithm, 
where $H$ is called the \emph{entropy}  and $H(\bm z, \hat{\bm z} )$ is the \emph{joint entropy} for cluster allocations. Intuitively, this distance is small if the amount of information two clusterings share, captured by the joint entropy, is close to the individual information of each clustering, captured by their entropy.

The posterior expectation of $L_\text{VI}$ by~\eqref{eq:loss_approx} involves $\mathbbm E \left[ \log_2  \left( n_{kk'}\right) \mid \bm X \right] $, computation of which requires the contingency counts for each sample. Instead, \cite{wade18} find a lower bound on the posterior expected loss, which only requires the computation of the posterior similarity matrix and leads to the loss function
\begin{equation}
L_{\text{VILB}}(\bm z,\hat{\bm z})    = \sum_{k'=1}^{K(\hat{\bm z})} n_{k'} \log_2 n_{k'} - 2 \sum_{i =1}^N \log_2 \left[ \sum_{j=1}^N \mathbbm 1\left(\hat z_i = \hat z_j \right) P_{ij} \right],
\end{equation}
where we omit terms that do not depend on $\hat{\bm  z}$. We use the lower bound approximation in our implementation.

% \subsection{Average silhouette}

\subsection{Optimisation}
\label{sec:opt}
Loss optimisation over all possible clusterings is infeasible for computational reasons. Indeed, if $\mathcal C (N)$ is the set of all clusterings of $\{1, \ldots, N\}$, then $|\mathcal C (N)|$ grows very rapidly as a function of $N$. The number of partitions into two clusters is $2^{N-1}-1$, so a simple lower bound is $2^{N-1}-1 \leq |\mathcal C (N)|$, indicating that optimisation of the loss function is NP-hard  and it is necessary to consider a reduced set of clusterings. We approximate the optimisation as
\begin{equation}
\label{def:opt}
     \bm z^* = \argmin_{\hat{\bm z} \in \mathcal C (N)} \mathbbm E \left[ L(\bm z,\hat{\bm z}) \mid \bm X \right]
         \approx \argmin_{\hat{\bm z} \in S} \mathbbm E \left[ L(\bm z,\hat{\bm z}) \mid \bm X \right],
\end{equation}
where $S \subset \mathcal C(N)$. In the following subsections we consider various choices of $S$.

We also mention more recent optimisation methods that we do not consider for brevity. \cite{rastelli18} use a greedy search algorithm and avoid computing the posterior similarity matrix, which is a costly intermediate step when $N$ is large, but nevertheless require certain types of loss functions which can be updated efficiently when a single $z_i$ is changed. \cite{dahl21} extend this optimisation algorithm with parallel programming and further search steps. However, these two approaches have not yet found wide use in practice, so we restrict ourselves to popular methods.

\subsubsection{Hierarchical clustering}

A common post-processing approach is to convert the posterior similarity matrix $\bm P$ into a metric between any two observations and then apply standard hierarchical clustering methods \citep{medvedovic04,fritsch09}. 
For this we define the posterior dissimilarity matrix $\bm D$ as $D_{ij} = 1 - P_{ij}$ for $i,j = 1, \ldots,N$, and perform agglomerative clustering on $\bm D$ with either complete or average linkage. 

Denoting the cluster allocation with $k$ clusters by $\hat{\bm z}(k)$, \cite{fritsch09} suggest optimising the loss function in \eqref{def:opt} with $S = \left\{ \hat{ \bm z}(k) : k = 1, \ldots, k_{\text{max}} \right\}$ for some upper bound $k_{\text{max}}$ such as $\lfloor N / 8 \rfloor$. If the posterior similarity matrix is already computed, this approach is very efficient, as we can update the linkage distances in constant time at every merging step, for a total of $O(N)$ operations. However, this yields only a single element of $S$ per value of $k$, whereas a desirable property of $S$ would be concentration around an appropriate numbers of clusters.

\cite{medvedovic04} introduced hierarchical clustering without loss function optimisation, instead obtaining $\bm z^*$ directly as the clustering that exceeds a linkage distance $1-\varepsilon$ across any pair of clusters. \cite{fritsch09} suggest  $1-\varepsilon = 0.99$, which we consider as a default. Unlike other methods we present, Medvedovic clustering does not depend on a loss function and does not seem to be expressible in decision-theoretic terms.

\subsubsection{Partitioning around medoids}
\cite{liverani15} suggest optimisation using the popular $k$-medoids algorithm, which generalises $k$-means for any metric. More precisely, they use $D_{ij} = 1 - P_{ij}$ as a distance metric and choose partitioning around medoids as an efficient implementation of $k$-medoids.   \cite{kaufman90} improve this approach by instead successively swapping medoids to minimize a global cost function. Nonetheless, when $N$ is large, partitioning around medoids is also inefficient as it requires $O\left[ k(N-k)^2 \right]$ iterations. We may therefore consider more efficient approaches such as that of \cite{schubert19}.

By denoting the output clustering of k-medoids with $k$ clusters by $\hat{\bm z}(k)$, \cite{liverani15} suggest optimising the loss function in \eqref{def:opt} with
$S = \left\{ \hat{\bm  z}(k) : k = 1, \ldots, k_{\text{max}} \right\}$ for some upper bound $k_{\text{max}}$. We choose $k_{\text{max}} = \lfloor N/8 \rfloor$ in our implementation.

\subsubsection{Sample search}
A straightforward choice for $S$ is the set of MCMC samples $S = \left\{ \bm z^{(1)}, \ldots , \bm z^{(M)} \right\}$ \citep{dahl06}. The resulting summary clustering is a posterior sample $\bm z^{(m)}$, which also provides associated sample component parameters $\bm \theta^{(m)}$. However, if all samples inherit an undesired property from the sampling method or the model itself,  such as overestimation of the number of components, the summary clustering will also have this property.

\section{Evaluation}

We now summarise the results of our analysis applying MFMs and DPMs, with and without conjugate hyperpriors on the component parameters, and with different summarisation methods.  We use three scenarios with simulated data and one with real data. 

The first scenario corresponds to a sample with a moderate sample size for which MFMs and DPMs are similar. Since the marginal posteriors for the number of clusters are highly concentrated at the truth, this easy case sets a baseline for comparing the different summarisation methods.

The second scenario corresponds to a larger sample size at which the DPM and MFM start to differ and which is more challenging for the summarisation methods. We focus on the DPM with hyperpriors since the corresponding posterior for the number of clusters is the flattest, with a mode that is higher than the truth in some of the simulated datasets. 

The third scenario is used to explore how a common type of misspecification affects summarisation methods, by modelling correlated data using multivariate Gaussian components with diagonal covariance matrices.

Finally, we show results for the gene expression dataset used in \citet{cai21}, where we expect to observe both the effects of DPM inconsistency and of misspecification. Again, we compare the outcomes of the different summarisation methods and use cell types as a proxy for true labels.

\label{chap:results}

 \subsection{Modelling specification}
We consider continuous data and multivariate normal mixture models. In the multivariate models we consider either a model with full covariance matrix or one with a diagonal covariance matrix. Both assumptions are common in practice, and in particular the restriction to a diagonal covariance has valued computational advantages.
The component distributions with a full covariance specification are
\begin{equation}
    \begin{aligned}
    \label{def:mvn}
    \mu_k  &\sim \mathcal{N}_p\left(\widehat{ \mu}, \widehat{ C} \right), \\
     \Sigma_k  &\sim \text{InverseWishart}\left(p,\widehat{C}\right), \\
    X_i \mid  z_i,  \bm \mu, \bm \Sigma &\sim \mathcal{N}_p\left(  \mu_{z_i}, \Sigma_{z_i} \right), \quad  i = 1, \ldots, N,
    \end{aligned}
\end{equation}
where $\widehat{ \mu} = N^{-1}\sum_{i=1}^N X_i$ is the sample mean, $\widehat{ C} = N^{-1}\sum_{i=1}^N (X_i - \widehat{ \mu}) (X_i - \widehat{ \mu})^T $ is the sample covariance and $k$ is the component index. As it is common to include hyperpriors to give models more flexibility or to incorporate prior assumptions, we also consider an alternative model with conjugate hyperpriors on the component parameters, i.e., 
\begin{equation}
    \begin{aligned}
    \label{def:mvnh}
    m_k &\sim \mathcal N_p\left(\widehat{\mu}, \widehat{C} \right),\\
    C_k &\sim  \text{InverseWishart}_p\left(p,\widehat{C}\right), \\
    \nu_k - p + 1 &\sim \text{Gamma}\left(\alpha = 2, \beta = 2\right),\\
    W_k &\sim \text{InverseWishart}_p\left(p,\widehat{ C}\right), \\
     \mu_k \mid m_k, C_k &\sim \mathcal{N}_p\left(m_k, C_k\right), \\
     \Sigma_k \mid \nu_k, W_k  &\sim \text{InverseWishart}_p\left(\nu_k,W_k\right), \\
    X_i \mid  z_i, \bm \mu, \bm \Sigma &\sim \mathcal{N}_p\left( \bm \mu_{z_i}, \bm \Sigma_{z_i} \right), \quad  i = 1, \ldots, N.
    \end{aligned}
\end{equation}

For the diagonal covariance model we standardize the observations beforehand, so our modelling assumptions can rely on standardised data. Hence, we have
\begin{equation}
    \begin{aligned}
    \label{def:mvnaac}
    \lambda_{kd} &\sim \text{Gamma}(1,1),\\ 
    \mu_{kd} \mid \lambda_{z_id}^{-1}  &\sim \mathcal{N}\left(0, \lambda_{z_id}^{-1}\right), \\
    X_{id} \mid  z_i, \bm \mu, \bm \lambda &\sim \mathcal{N}\left( \mu_{z_id}, \lambda_{z_id}^{-1}\right),  \quad i = 1, \ldots, N,
    \end{aligned}
\end{equation}
for each dimension $d=1,\ldots,p$ independently. 

We consider two types of allocation prior, either a mixture of finite mixtures \eqref{def:mfm} or a Dirichlet process mixture \eqref{def:dpm}. For the first, $\bm{\pi} \mid K \sim \text{Dirichlet}_K(1,\ldots,1)$ and $K \sim \text{Geometric}(0.1)$, following \cite{miller18}. For the second, $\alpha \sim \text{Exponential}(1)$. 

For both models we obtain MCMC samples with the split-merge sample with four independent chains, each with $2000$ iterations, and the $100$ first iterations removed as burn-in. We thin the MCMC samples by keeping every second iteration, giving $M = 3800$ samples in total. \cite{miller18} show fast convergence of the split-merge sampler, so we choose the sample size sufficiently large to be representative while limiting the running time. We assess convergence using the Geweke and the Gelman--Rubin $\hat R$ diagnostics  across multiple chains \citep{geweke91,gelman92,vehtari20}.

\subsection{Simulation study}

In order that the true number of components is known, we generate synthetic data using a finite mixture model from \cite{miller18} with four clusters. Our synthetic bivariate observations are distributed as
\begin{equation}\label{eq:sim-data}
    \begin{aligned}
    z_i &\sim \text{Categorical}( \bm \pi ), \quad i=1, \ldots,N, \\
    X_i &\sim \mathcal N_2 \left(\mu_{z_i}, \Sigma_{z_i}\right), \quad i=1, \ldots,N,
    \end{aligned}    
\end{equation}
where
\begin{equation}
\begin{gathered}
    \bm \pi = (0.44, 0.3, 0.25,0.01),\\[1pt]
    \mu_1 = (4,4)^T, \, \mu_2 = (7,4)^T, \, \mu_3 = (6,2)^T, \, \mu_4 = (8,11)^T, \\[1pt]
    \Sigma_1 = \begin{pmatrix}
    2 & 0\\
    0 & 2
    \end{pmatrix},\, \Sigma_2 = R \begin{pmatrix}
    2 & 0\\
    0 & 2
    \end{pmatrix}R^T, \, R = \begin{pmatrix}
    \cos(\pi/4) & -\sin(\pi/4)\\
    \sin(\pi/4) & \cos(\pi/4)
    \end{pmatrix},\\[1pt]
    \Sigma_3 = \begin{pmatrix}
    3 & 0\\
    0 & 0.1
    \end{pmatrix}, \, 
    \Sigma_4 = \begin{pmatrix}
    0.1 & 0\\
    0 & 0.1
    \end{pmatrix}.
\end{gathered}    
\end{equation}

\subsubsection{Moderate sample size}
\label{subsec:example}
%\subsubsection*{Illustrative synthetic dataset}

\begin{figure}[!h!]
     \centering
     \begin{subfigure}[b]{\textwidth}
         \centering
    \includegraphics[width=0.8\linewidth]{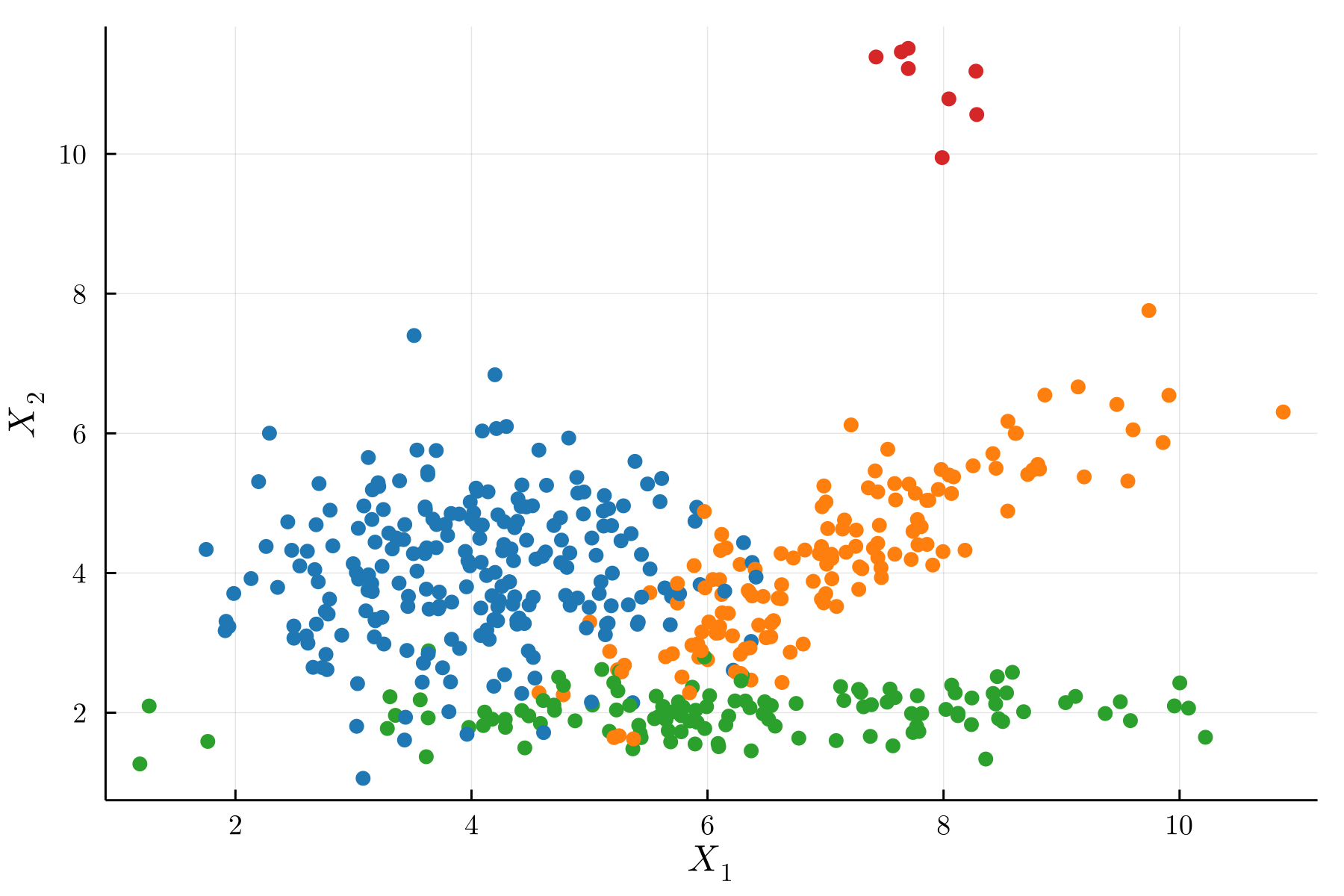}
        %  \caption{}\label{fig:example_data_a}
     \end{subfigure}

    %  \begin{subfigure}[b]{0.49\textwidth}
    %      \centering
    %     \includegraphics[height=2.3in]{figures/example/example_samples1}
    %      \caption{}
    %      \label{fig:example_data_b}
    %  \end{subfigure}
    %  \hfill
    %  \begin{subfigure}[b]{0.49\textwidth}
    %      \centering
    %      \includegraphics[height=2.3in]{figures/example/example_samples2}
    %      \caption{}
    %      \label{fig:example_data_c}
    %  \end{subfigure}
        \caption{Synthetic Gaussian data with four clusters and $N = 500$ observations.}
        \label{fig:example_data}
\end{figure}

To obtain representative results we generated $50$ independent datasets with $N= 500$ observations from model \eqref{eq:sim-data}. Figure~\ref{fig:example_data} shows one of these datasets, which display overlap between the component densities. Uncertainty between clusters is often seen in applications, so by fixing $\bm \pi$, $\bm \mu$ and $\bm \Sigma$ we control the difficulty of the clustering task. If instead we generate data from the model with hyperpriors, then we do not control distances between component means, which may confound our interpretation of results.

% \begin{figure}
%     \centering
%     \includegraphics[width=0.8\linewidth]{figures/example/example_data.png}
    
%         \includegraphics[width=0.9\linewidth]{figures/example/example_samples.png}
%     \caption{Synthetic Gaussian data with four clusters and $N = 500$ observations. Colors indicate the true generating cluster allocations.}
%     \label{fig:example_data}
% \end{figure}

 %Figures~\ref{fig:example_data_b} and \ref{fig:example_data_c} gives two representative examples of allocations sampled from the posterior when the MFM model is fitted to the data shown in Figure~\ref{fig:example_data}. Due to label-switching the particular values $z_i$, here represented by different colours, are arbitrary, but from visual inspection we recognise the true cluster allocation with some movement in the overlapping regions and more uncertainty for low density observations.

% \begin{figure}
%     \centering
%     \includegraphics[width=0.9\linewidth]{figures/example/example_samples.png}
%     \caption{Two sample allocations for the mixture of finite mixtures (MFM) model on synthetic data with $N = 500$, coloured by cluster. Differences in colour between samples are due to label-switching.}
%     \label{fig:example_samples}
% \end{figure}

We perform estimation with the full covariance models \eqref{def:mvn} and \eqref{def:mvnh} and four model types: the mixture of finite mixtures with and without a hyperprior on the parameters of the mixture components and the Dirichlet process mixture with and without hyperpriors, which we respectively abbreviate to MFMH, MFM, DPMH and DPM. At each iteration we obtain a sample of component allocations $\bm z$ and the number of active components $T$.

\begin{figure}
    \flushleft
    \begin{subfigure}{0.49\linewidth}
        \includegraphics[width=1\linewidth]{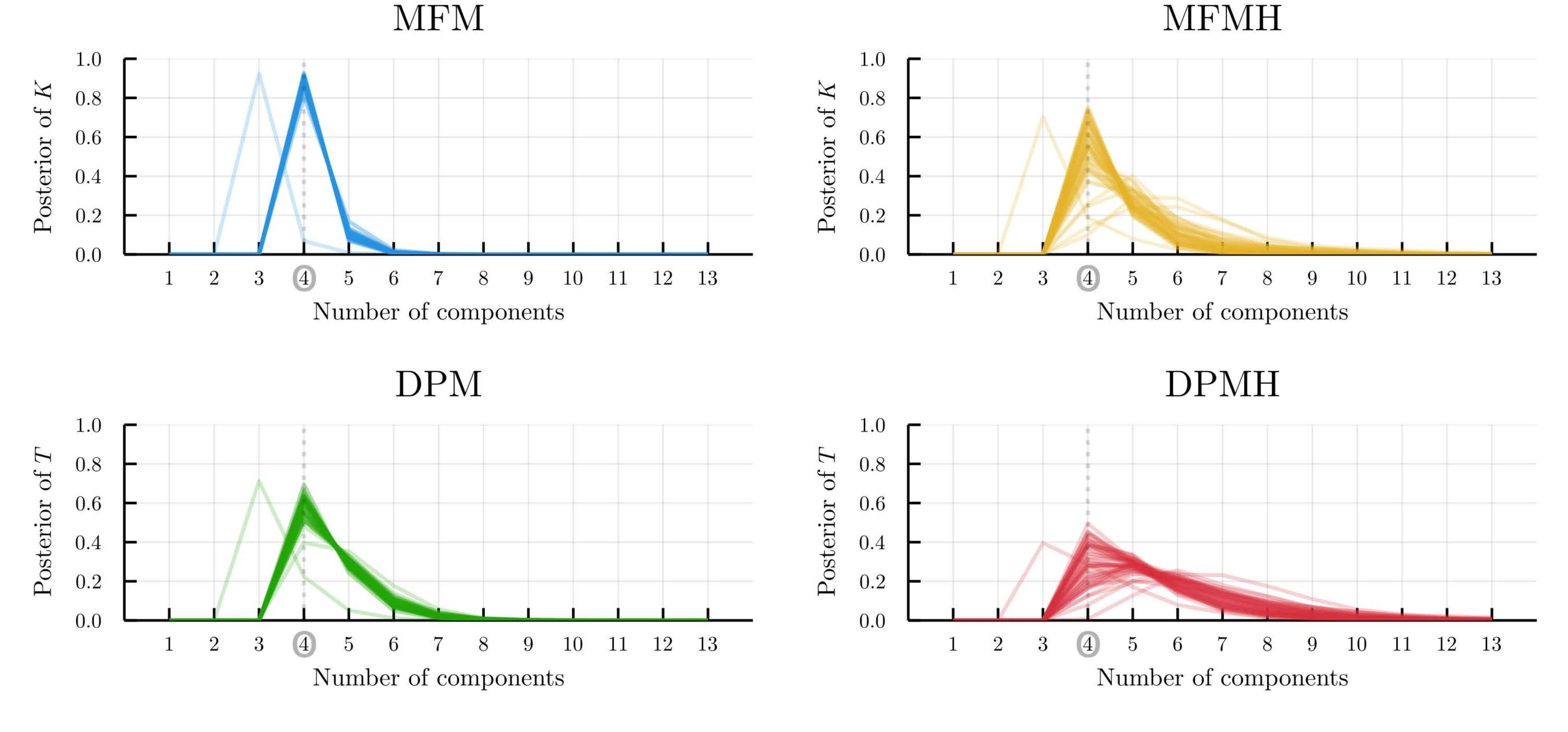}
        \subcaption{MFM}
    \end{subfigure}
    \hfill
    \begin{subfigure}{0.49\linewidth}
        \includegraphics[width=1\linewidth]{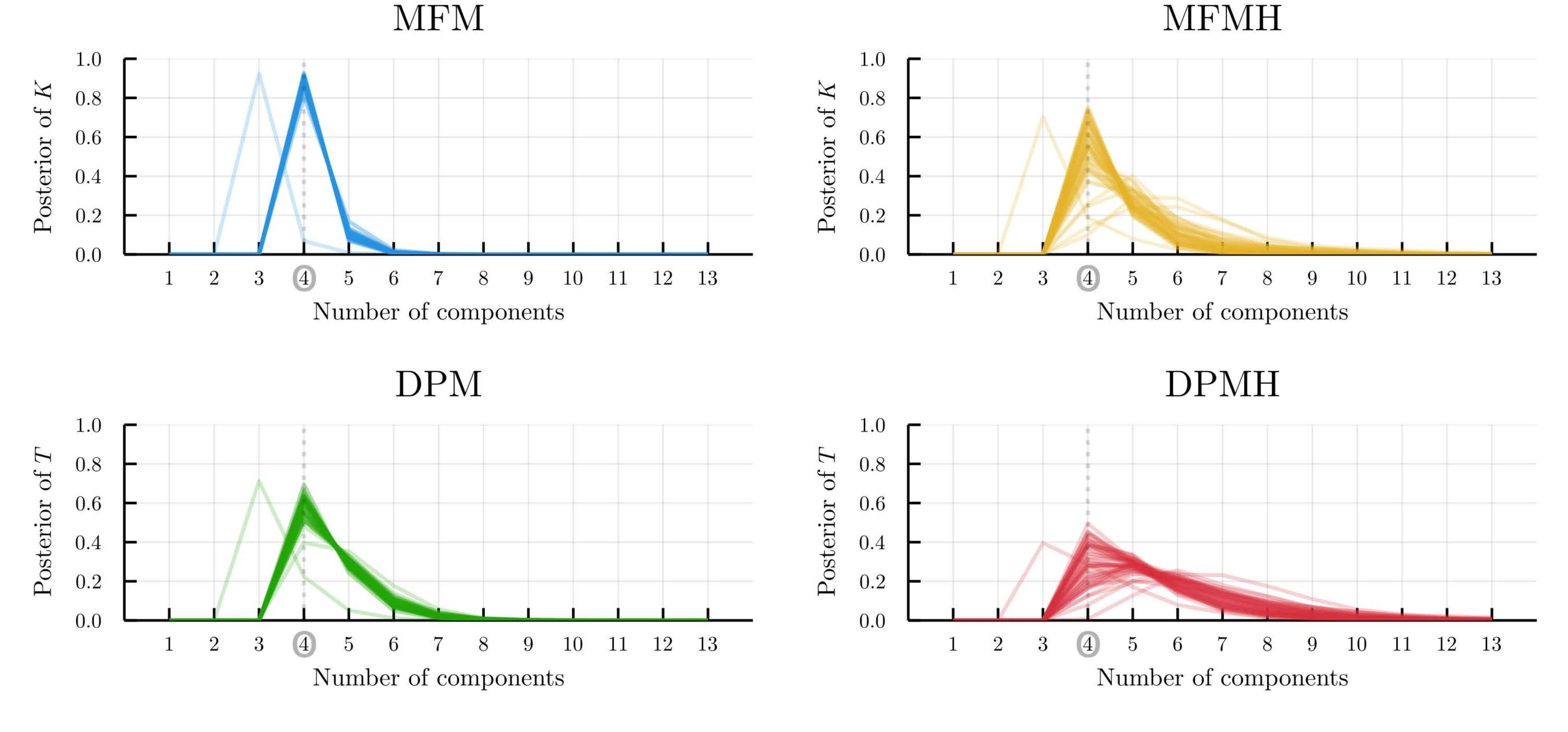}
        \subcaption{MFMH}
    \end{subfigure}
    
     \begin{subfigure}{0.49\linewidth}
        \includegraphics[width=1\linewidth]{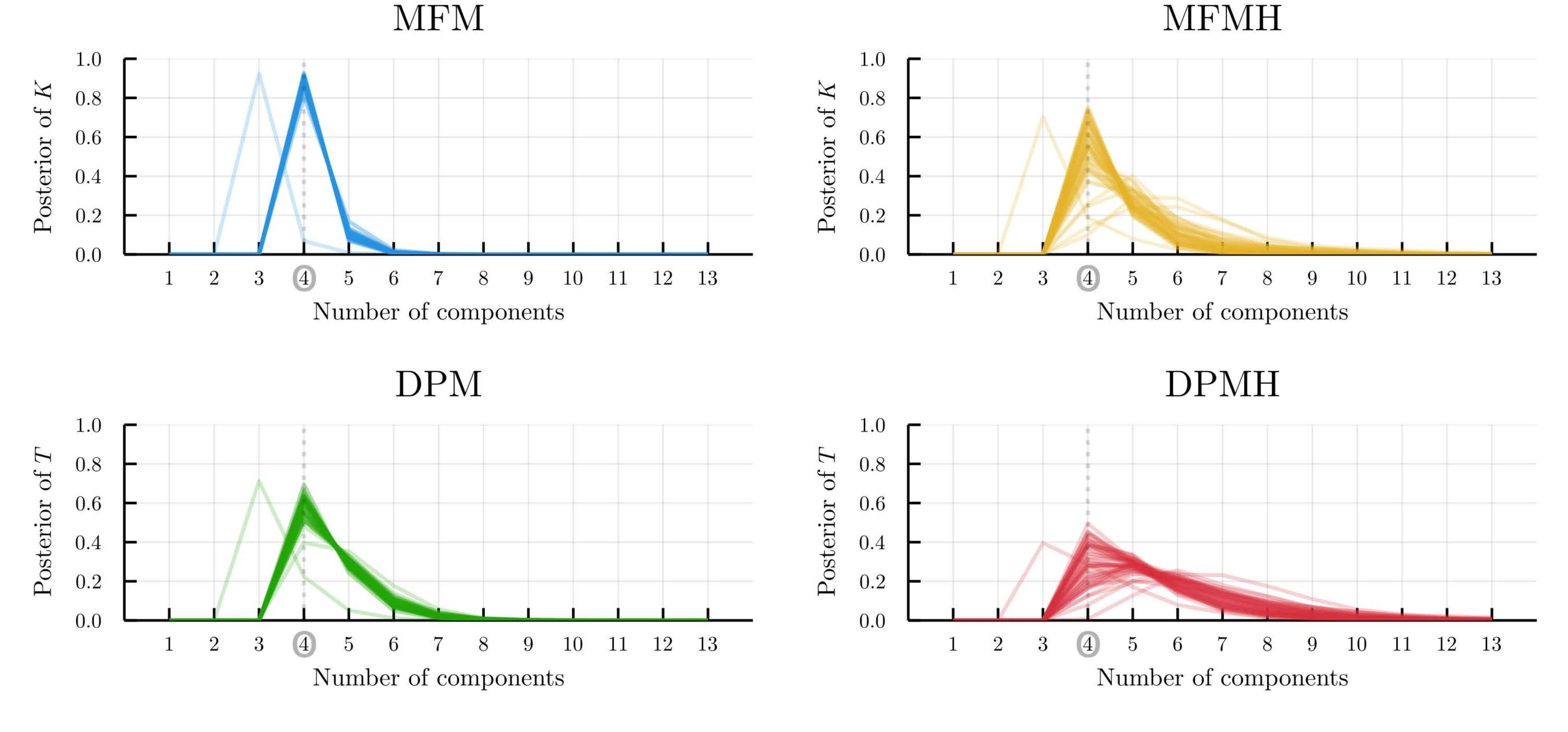}
        \subcaption{DPM}
    \end{subfigure}
    \hfill
    \begin{subfigure}{0.49\linewidth}
        \includegraphics[width=1\linewidth]{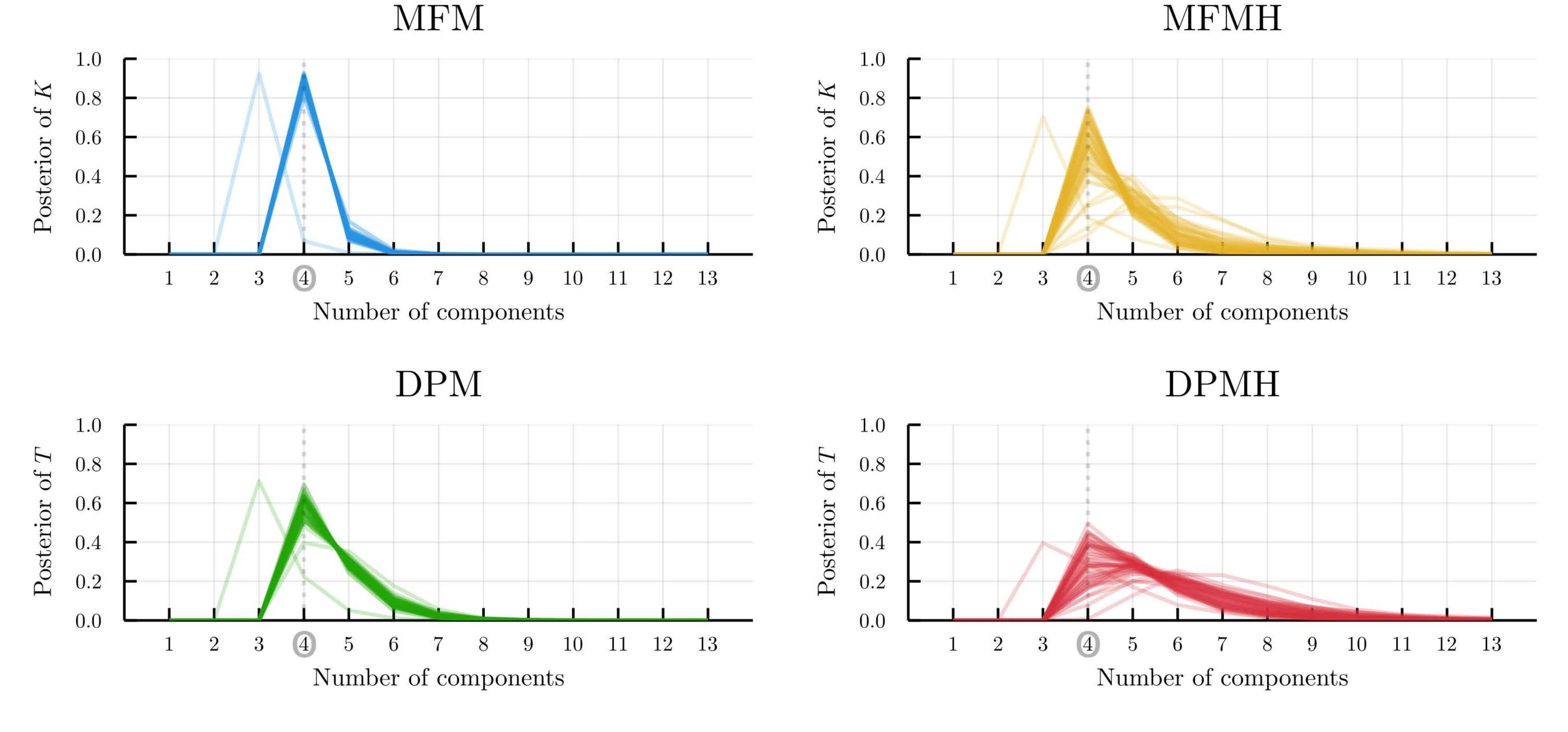}
        \subcaption{DPMH}
    \end{subfigure}
  
    \caption{Posteriors on the number of components for a mixture of finite mixtures (a) without and (b) with a hyperprior,  and for a Dirichlet process mixture (c) without and (d) with a hyperprior, for $50$ synthetic datasets with $N = 500$. There are in fact four components.}
    \label{fig:example_posts}
\end{figure}

Figure~\ref{fig:example_posts} shows every resulting posterior on the number of components. For clarity, we maintain a consistent colouring of the four model types in all following figures. To obtain the posterior of $K$ for the mixture of finite mixtures model, we use the posterior on $K$ given the number of active components $T$, provided by \cite{miller18}. Although the dataset in Figure~\ref{fig:example_data} is well-specified by the full covariance model and $N = 500$ is a moderate sample size, there are differences between the model types: the hyperpriors seem to flatten the posterior number of components towards higher values, and  the Dirichlet process mixture allows exploration of higher numbers of components during MCMC sampling.

The identifiability issues make diagnostics for MCMC samples of $\bm \theta$, $\bm z$ and $\bm \pi$ challenging, and we can only consider the number of components or $\alpha$. Figure~\ref{fig:example_trace} in the Appendix gives an example of the number of components for a single chain before thinning and including the burn-in samples when fitting each of our models to the data shown in Figure~\ref{fig:example_data}. %As expected, the models without hyperpriors converge more quickly to what seems to be their stationary distribution, possibly because of data dependent priors.

\begin{figure}
    \centering
    \includegraphics[width=1\linewidth]{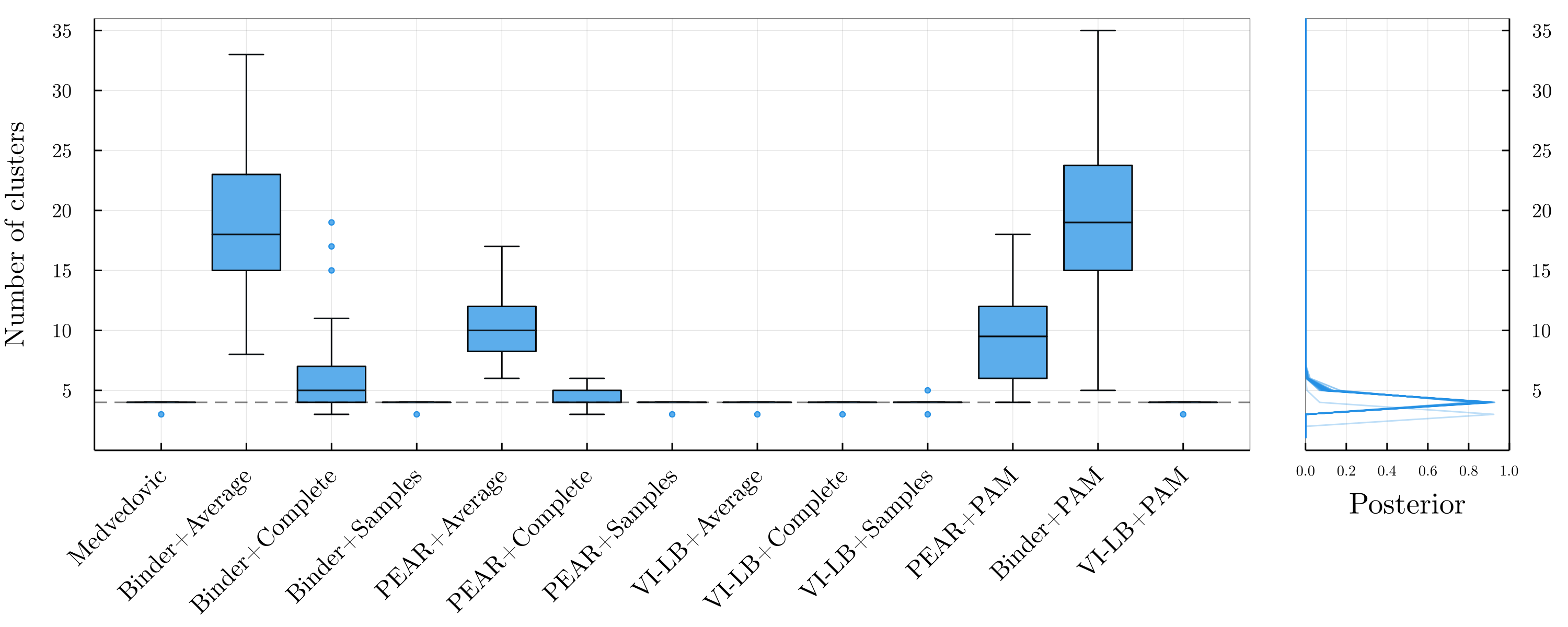}
    \caption{Number of clusters of summary clustering methods and posterior number of components for the mixture of finite mixtures (MFM) model without a hyperprior, for $50$ independent synthetic datasets with $N=500$. Summary names are composed of the loss function and the optimisation method, except for Medvedovic clustering.  The true number of components, four, is shown by the horizontal dashed line.}
    \label{fig:example_nclust}
\end{figure}

For every dataset and model type we derive the summary clusterings described in Section~\ref{chap:clust}, each of which gives a number of clusters.
As an example, Figure~\ref{fig:example_nclust} summarises the results for the mixture of finite mixtures model without a hyperprior. The posteriors at the right are those in Figure~\ref{fig:example_posts}. The summarisation methods with a loss function are Binder's loss (\emph{Binder}), the posterior expected adjusted Rand index (\emph{PEAR}), the lower bound of the variation of information (\emph{VI-LB}), average and complete linkage hierarchical clustering (\emph{Average} and \emph{Complete}), sample search (\emph{Samples}) and partitioning around medoids (\emph{PAM}). Although this case is rather simple, some methods vastly overestimate the number of clusters.

Our main focus is the number of clusters, but  this is linked to the underlying cluster allocations.
Nonetheless, an accurate clustering can coexist with an inaccurate number of clusters, at least when accuracy is measured with popular scores such as the adjusted Rand index.
The addition of small clusters has a minor impact on these scores provided the large clusters are correctly identified, as illustrated in the Appendix; see Figure~\ref{fig:example_ari}.

\subsubsection{Dirichlet process mixture inconsistency}
\label{subsec:res_dpm}

\begin{figure}
    \flushleft

        \centering
    \begin{subfigure}{0.49\linewidth}
        \includegraphics[width=1\linewidth]{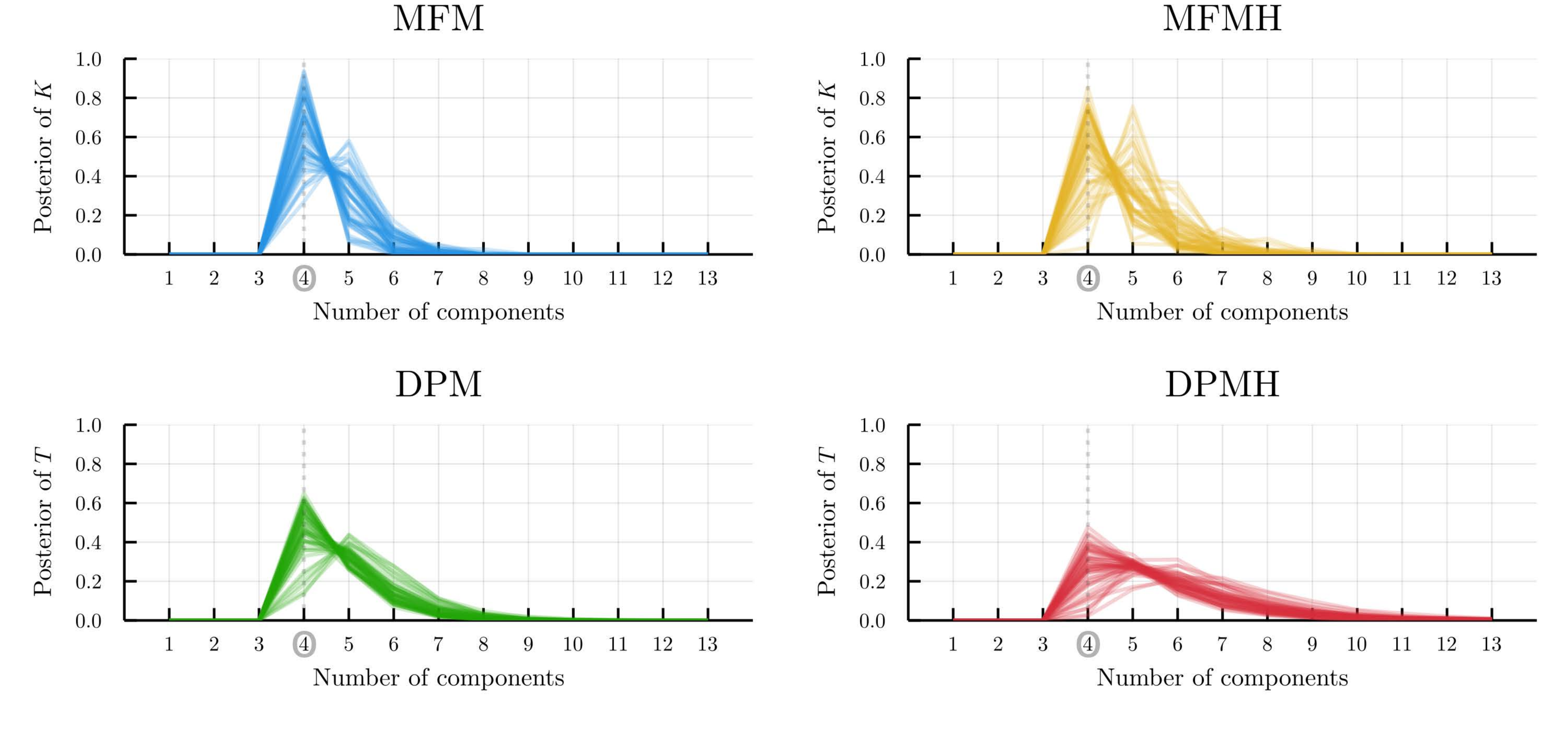}
        \subcaption{MFM}
    \end{subfigure}
    \hfill
    \begin{subfigure}{0.49\linewidth}
        \includegraphics[width=1\linewidth]{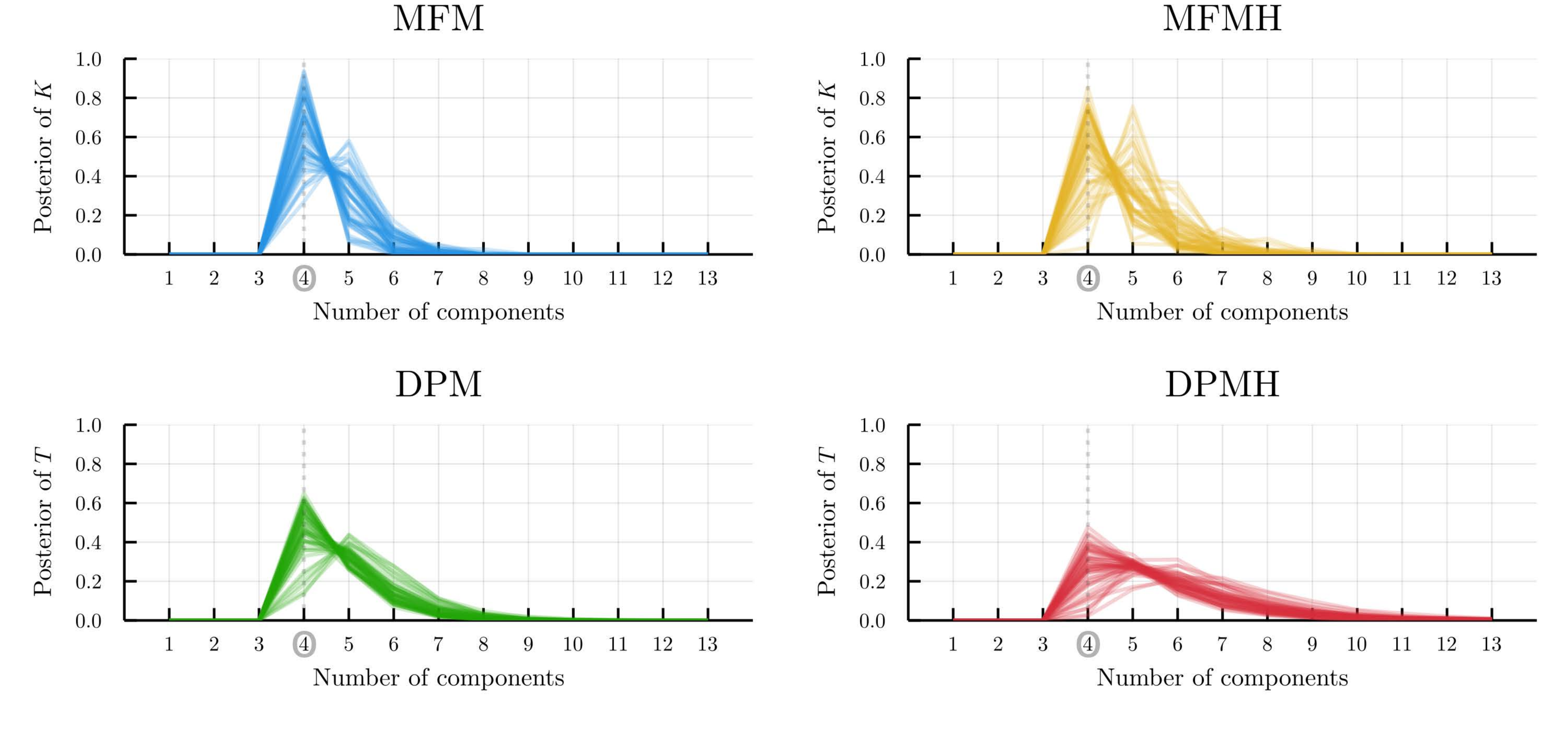}
        \subcaption{MFMH}
    \end{subfigure}
    
     \begin{subfigure}{0.49\linewidth}
        \includegraphics[width=1\linewidth]{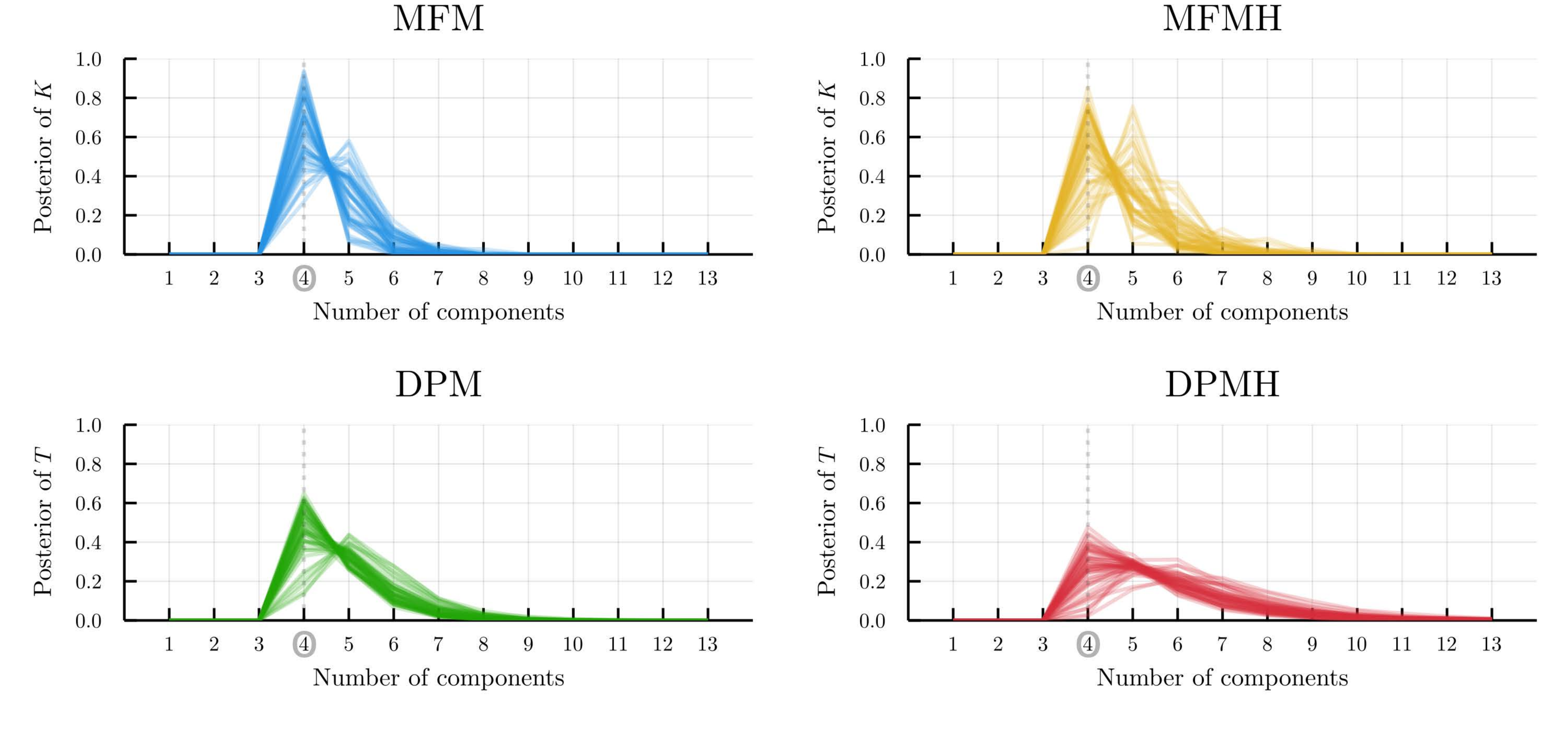}
        \subcaption{DPM}
    \end{subfigure}
    \hfill
    \begin{subfigure}{0.49\linewidth}
        \includegraphics[width=1\linewidth]{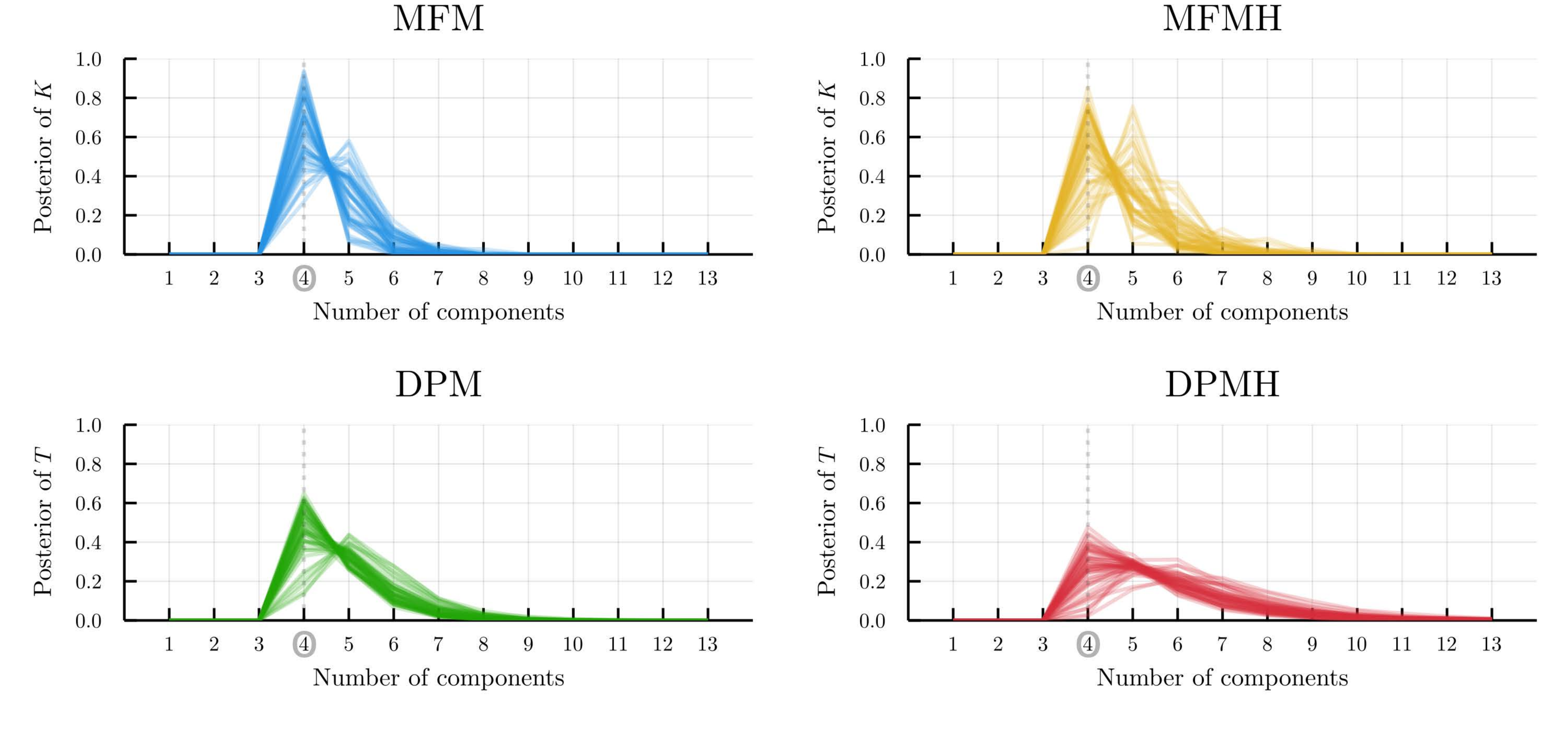}
        \subcaption{DPMH}
    \end{subfigure}

    \caption{Posteriors on the number of components for a mixture of finite mixtures (a) without and (b) with a hyperprior, and for a Dirichlet process mixture (c) without and (d) with a hyperprior, for $50$ synthetic datasets with $N = 10^4$. There are in fact four components.}
    \label{fig:dpm_post}
\end{figure}

To investigate the Dirichlet process mixture and asymptotic results of Section~\ref{subsec:dpm_incons} we set $N = 10^4$. We consider the Dirichlet process mixture with hyperpriors, as it shows the highest overestimation of the number of components even when $N = 500$, and in Figure~\ref{fig:dpm_post} show the posterior number of components for $50$ independent datasets.  The Dirichlet process mixture model tends to overestimate the number of components, in particular compared to the mixture of finite mixtures model.

The Dirichlet process mixture model tends to introduce small extra clusters, as noticed by \cite{miller14} and illustrated in Figure~\ref{fig:dpm_samp} in the Appendix. 

Figure~\ref{fig:dpm_nclust} gives the number of clusters for a selection of summary clusterings. We exclude results for Binder's loss and PEAR loss with average linkage, which can both massively overestimate the number of clusters, and we do not use partitioning around medoids because of its much longer running time when $N$ is large.
As expected, optimisation with sample search tends to follow the posterior distribution and leads to overestimation of the number of clusters, whereas Medvedovic clustering and variation of information both give the correct number of clusters for all $50$ datasets.  This confirms that summary clustering methods may eliminate any small extra clusters introduced by the DPM model. Indeed, this holds even for data whose posterior mode is appreciably larger than four and the posterior credibility of four components is low.

Figure~\ref{fig:dpm_clust} of the Appendix gives two examples of summary clusterings; PEAR loss with average linkage is representative of methods not included in Figure~\ref{fig:dpm_nclust}, as it gives $161$ clusters, while variation of information leads to four clusters only.
Both Binder's loss and PEAR loss with complete linkage optimisation tend to create small clusters in the overlaps between components.
This occurs independently of the model type, so it is a different phenomenon from Dirichlet process mixture inconsistency; indeed, \cite{wade18} already notice this for Binder's loss.
Our results in Section~\ref{subsec:example} show that Binder's loss leads to poor results for the number of clusters that for $N = 500$.
%[See Appendix ... for more details on Binder's loss overestimation of the number of clusters].
For PEAR loss we could investigate whether the approximation by \cite{fritsch09} is the reason for these additional clusters.  Variation of information successfully adjusts for small extra clusters introduced during sampling, but the resulting clustering differs from the generating component allocation, as the clusters do not have Gaussian shapes.
We conjecture that the borders between clusters approximate the contour lines of equal component densities, e.g., $\{ x: f(x \mid \theta_1) = f(x \mid \theta_2) \}$, which would explain the straight boundaries between clusters.

%  Mov to supplement

%  Move to supplement
% \begin{figure}
%     \centering
%     \includegraphics[width=0.8\linewidth]{figures/dpm/dpm_compsizes.png}
%     \caption{Average posterior component proportions for the Dirichlet process mixture (DPMH) model with a hyperprior over $50$ independent synthetic datasets with $N= 10^4$ observations. Proportions are reordered for every MCMC sample to avoid non-identifiability. Cross-marks give the four true generating mixture proportions.}
%     \label{fig:dpm_compsizes}
% \end{figure}

% \begin{figure}
%     \centering
%     \includegraphics[width=0.6\linewidth]{figures/dpm/dpm_psm.png}
%     \caption{Posterior similarity matrix for the Dirichlet process mixture  model with a hyperprior (DPMH) on synthetic data with $N = 10^4$. Observations are reordered with hierarchical clustering for visibility.}
%     \label{fig:dpm_psm}
% \end{figure}

\begin{figure}
    \centering
    \includegraphics[width=1\linewidth]{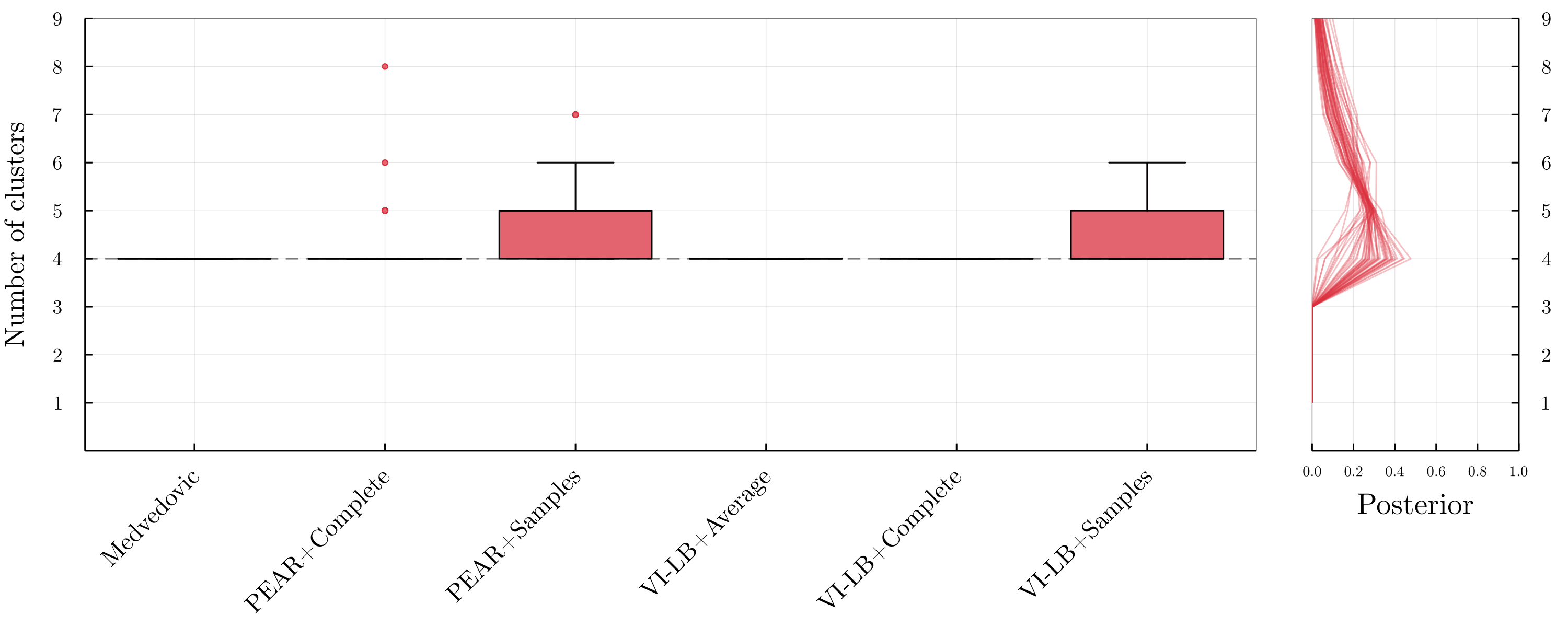}
    \caption{Number of clusters of summary clustering methods and posterior number of components for the Dirichlet process mixture model with a hyperprior (DPMH), for $50$ independent synthetic datasets with $N=10^4$. Summary names are composed of the loss function and the optimisation method, except for Medvedovic clustering.}
    \label{fig:dpm_nclust}
\end{figure}

%  Move to supplement

\subsubsection{Misspecification inconsistency}
\label{subsec:res_miss}
We now model the previous correlated data with $N = 10^4$ using the diagonal covariance model \eqref{def:mvnaac}, which is popular for computational reasons, to investigate asymptotic results from Section~\ref{subsec:miss}. The model~\eqref{def:mvnaac} is misspecified, because it cannot express the correlation from $\Sigma_2$. Indeed, with this choice we can specify a conjugate prior on the component distributions and sample directly from the component prior, thereby avoiding the conditional Gibbs update during the split-merge algorithm. Moreover the number of covariance parameters reduces from $p(p+1)/2$ to $p$ for every component, so sampling should be more stable in high-dimensional cases. %A diagonal covariance is sometimes assumed even though some correlation within observations is expected. 

\begin{figure}
    \centering
           \centering
    \begin{subfigure}{0.49\linewidth}
        \includegraphics[width=1\linewidth]{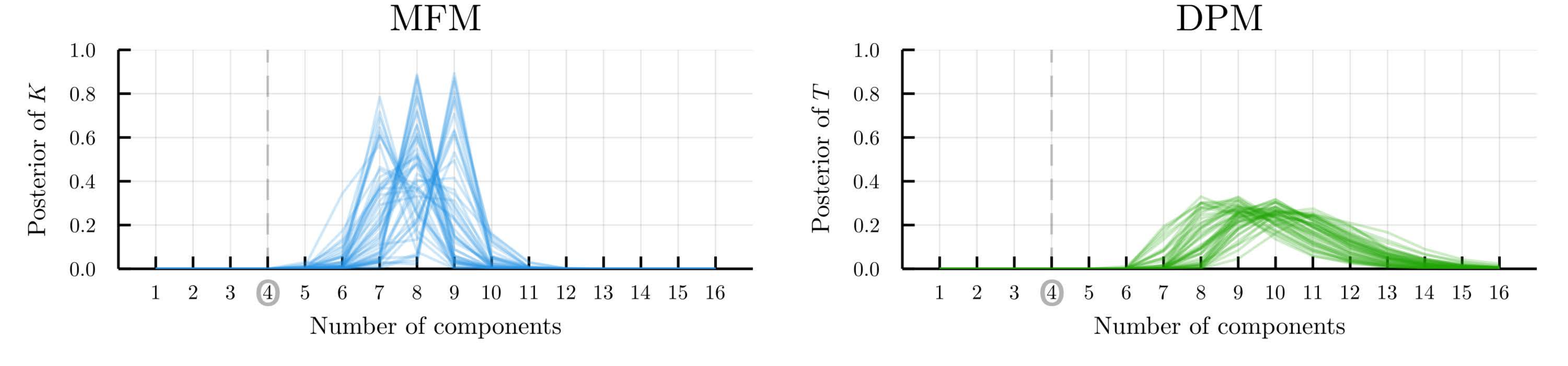}
        \subcaption{MFM}
    \end{subfigure}
    \hfill
    \begin{subfigure}{0.49\linewidth}
        \includegraphics[width=1\linewidth]{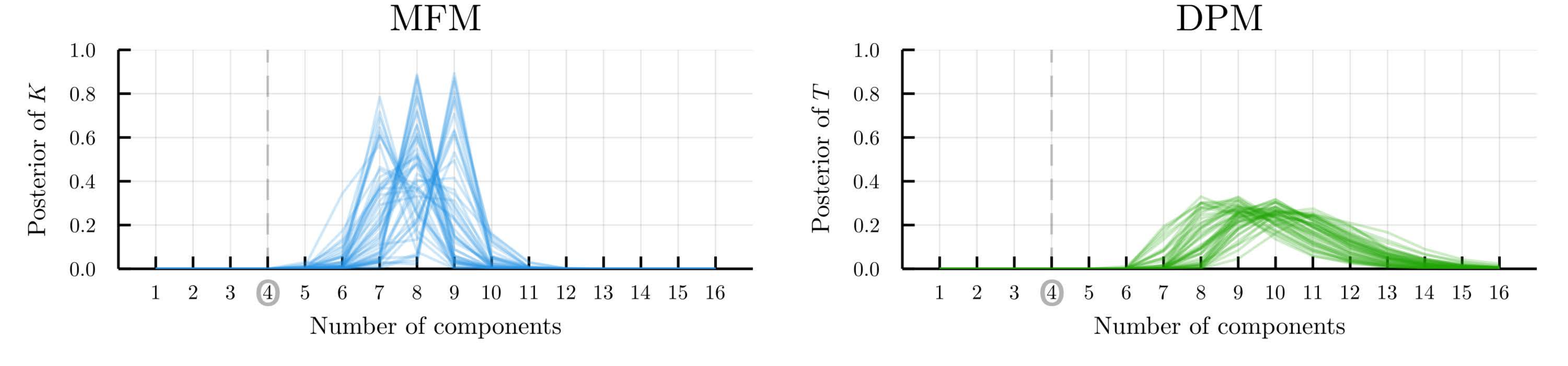}
        \subcaption{DPM}
    \end{subfigure}

    \caption{Posterior on the number of components under misspecification for (a) the mixture of finite mixtures; and (b) Dirichlet process mixture, with diagonal covariance on $50$ independent synthetic datasets with $N= 10^4$.  There are in fact four components.}
    \label{fig:miss_post}
\end{figure}

We again generate $50$ datasets of correlated Gaussian mixture observations and sample from the diagonal covariance model~\eqref{def:mvnaac}. Figure~\ref{fig:miss_post} shows   the posterior numbesr of components for two model types we consider. The MFM model gives concentrated posterior distributions for the number of components that contract onto a single value, though this varies across datasets. It is unclear why the model is sensitive to data generation, as we would expect almost identical data when $N$ is so large. MCMC samples, not shown here, show the creation of nearly identical overlapping components sharing observations of the well-specified components. One reason for this could be that, as $K$ gets large, the symmetric Dirichlet prior attributes more density to allocations with similar proportions, thereby giving potentially lower posterior credibility to uneven proportions. To investigate this we could choose a smaller $\gamma$, such as $\gamma = 1/K$. The DPM model posterior distributions also vary across datasets, but are less concentrated than those for the MFM model. This difference may arise because of the overestimation described in Section~\ref{subsec:res_dpm}.

\begin{figure}
\begin{subfigure}{1\linewidth}
    \centering
    \includegraphics[width=1\linewidth]{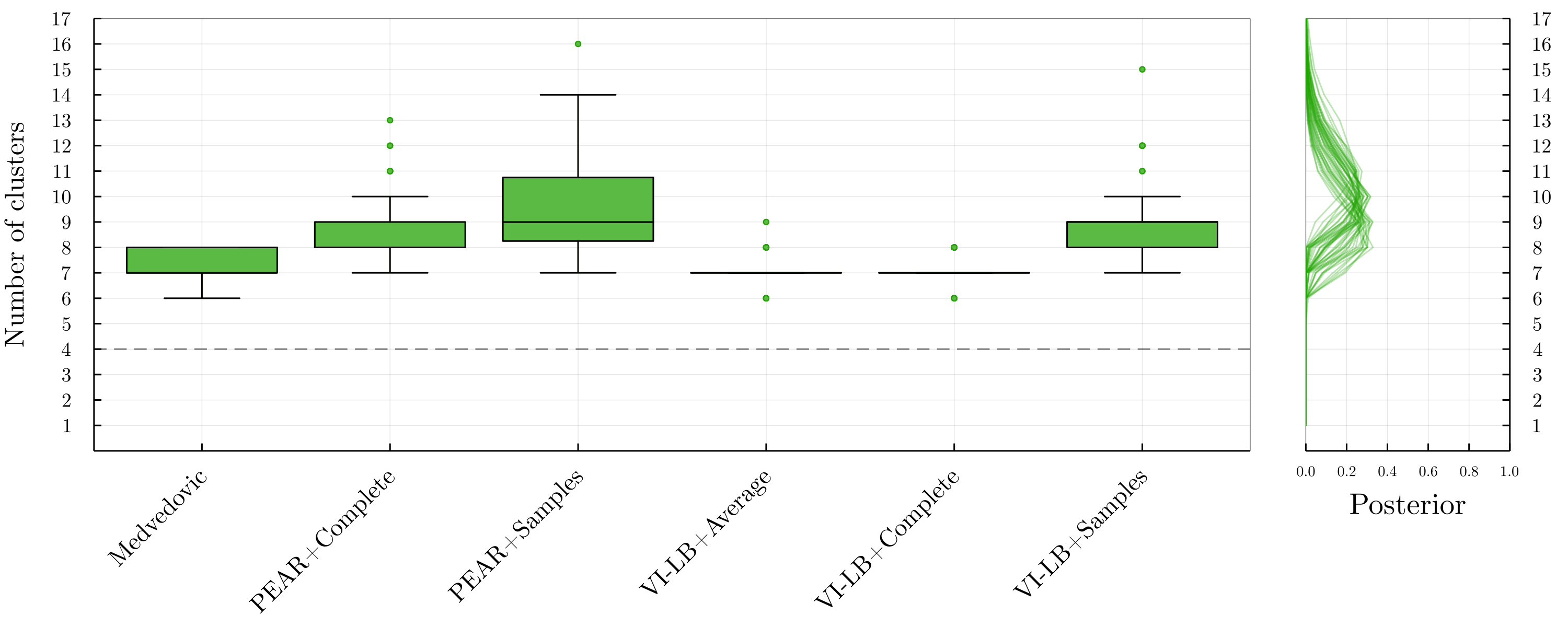}
    \caption{Dirichlet process mixture model.}
\end{subfigure}

\vspace{1cm}

\begin{subfigure}{1\linewidth}
    \centering
    \includegraphics[width=1\linewidth]{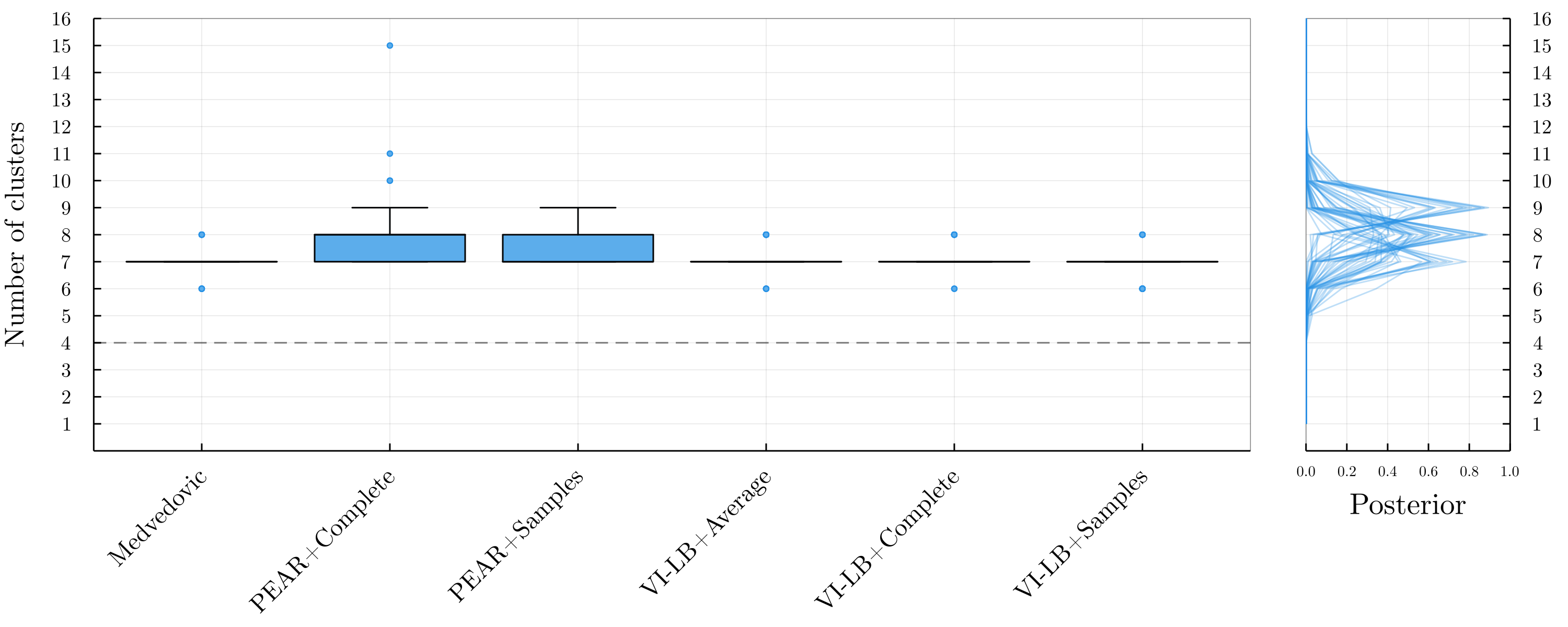}
    \caption{Mixture of finite mixtures model.}
\end{subfigure}
    \caption{Number of clusters of summary clustering methods and posterior number of components for (a) the misspecified mixture of finite mixtures (MFM) model and (b) the misspecified Dirichlet process mixture (DPM) model, for $50$ synthetic datasets with $N=10^4$. Summary names are composed of the loss function and the optimisation method, except for Medvedovic clustering.  The true number of components, four, is shown by the horizontal dashed line.}\label{fig:miss_nclust}
\end{figure}

Figure~\ref{fig:miss_nclust} shows the numbers of clusters for summarisation methods for the MFM and DPM models.
As in the previous section, we omit methods with significant overestimation. 
Overall summary clusterings seem to minimise the number of components compared to the posterior but they do not recover the true number of components.
Among these methods, variation of information leads to the most consistent reduction, giving seven clusters in most cases independently of the model type.
Sample search optimisation seems to follow the posterior distribution. The adjusted Rand index of summary clusterings, not shown here, is similar for all methods except for the slightly less accurate sample search, similarly to Figure~\ref{fig:example_ari}.

\begin{figure}
    \centering
    \begin{subfigure}{0.49\linewidth}
        \includegraphics[width=1\linewidth]{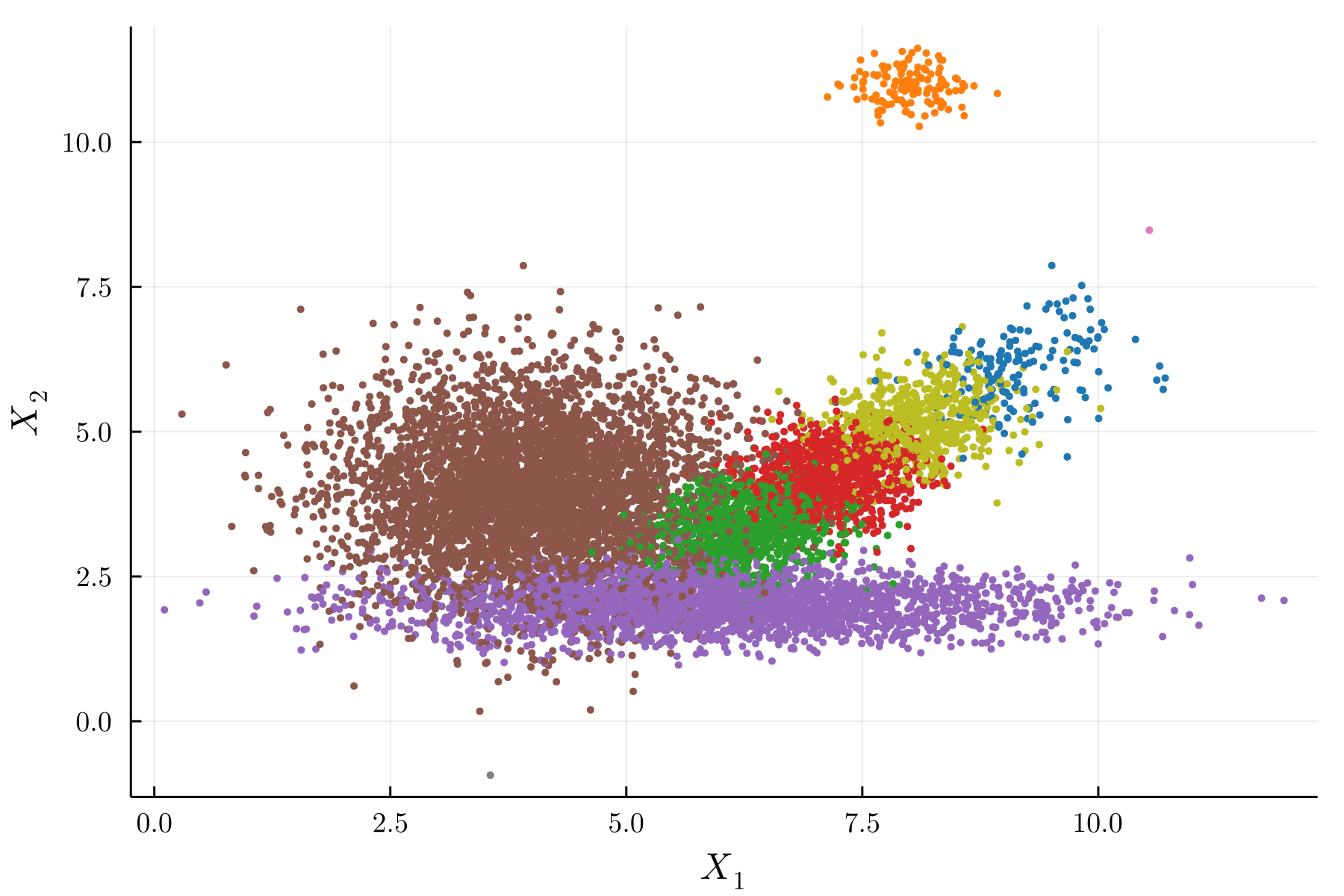}
        \subcaption{}
    \end{subfigure}
    \hfill
    \begin{subfigure}{0.49\linewidth}
        \includegraphics[width=1\linewidth]{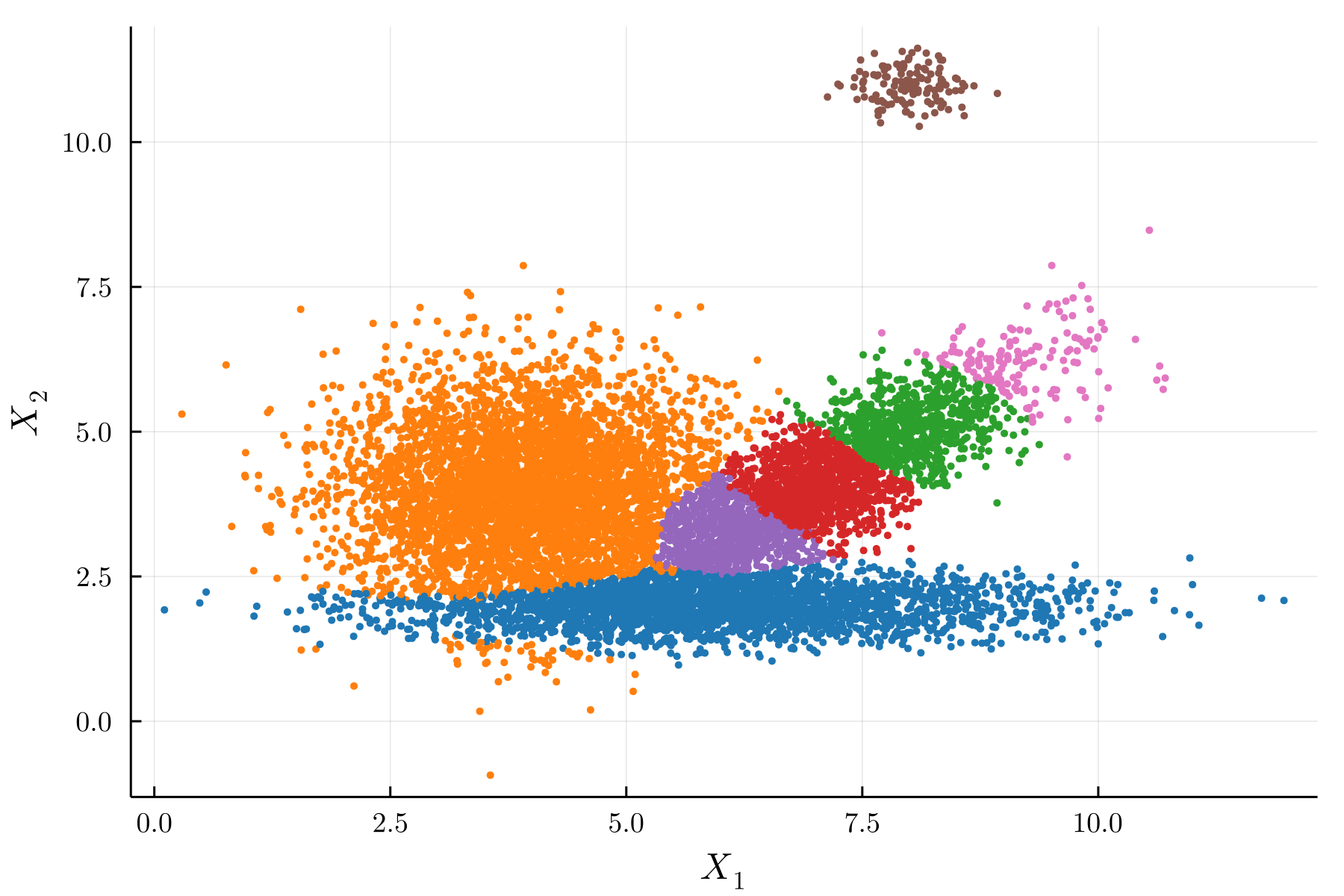}
        \subcaption{}
    \end{subfigure}
    \caption{Summary clusterings with variation of information optimised with (a) sample search and (b) complete linkage hierarchical clustering for the misspecified mixture of finite mixtures (MFM) MCMC samples on synthetic data with $N=10^4$.}
    \label{fig:miss_clust}
\end{figure}

The only component that cannot be expressed using the diagonal covariance model is the second component.  As $\Sigma_2$ is non-diagonal, the misspecified model introduces additional components. Figure~\ref{fig:miss_clust} shows an MCMC sample selected by the variation of information loss in which four components approximate the second component. The summary clustering with hierarchical clustering optimisation inherits the three additional components from the MCMC samples. It seems that summary clusterings do not adjust for misspecification in this example, but only avoid small additional clusters, in particular for the DPM model. %Figure~\ref{fig:miss_compsizes} gives the average posterior proportions and shows that compared to the previous section the misspecification leads to bigger additional clusters and we cannot visually recognise the true number of components.

\subsection{Gene expression data}

We now consider gene expression data from single-cell RNA sequencing of somatosensory cells \citep{zeisel15}, which have $N = 3005$ to address asymptotic theory, and which were also used by \cite{cai21} to illustrate the inconsistency result presented in Section~\ref{subsec:miss}.
The data are 19,972 gene measurements from 3005 cells with seven cell types. We pre-process the data following \cite{prabhakaran16} and use the $p=10$ genes with the largest standard deviations because we aim to use our full covariance model in addition to the the diagonal covariance model used by \cite{cai21}; \cite{prabhakaran16} took $p=558$. 
Modelling using the full covariance model is much slower for large $p$ due to the additional $p(p-1)/2$ correlation parameters, but sampling from both allows us to compare the diagonal covariance and more expressive full covariance models, though even the latter can be expected to be misspecified.   According to \cite{zeisel15}, there are seven known cell types, which we will consider as ground truth.

\begin{figure}
    \centering
    \begin{subfigure}{0.49\linewidth}
        \includegraphics[width=1\linewidth]{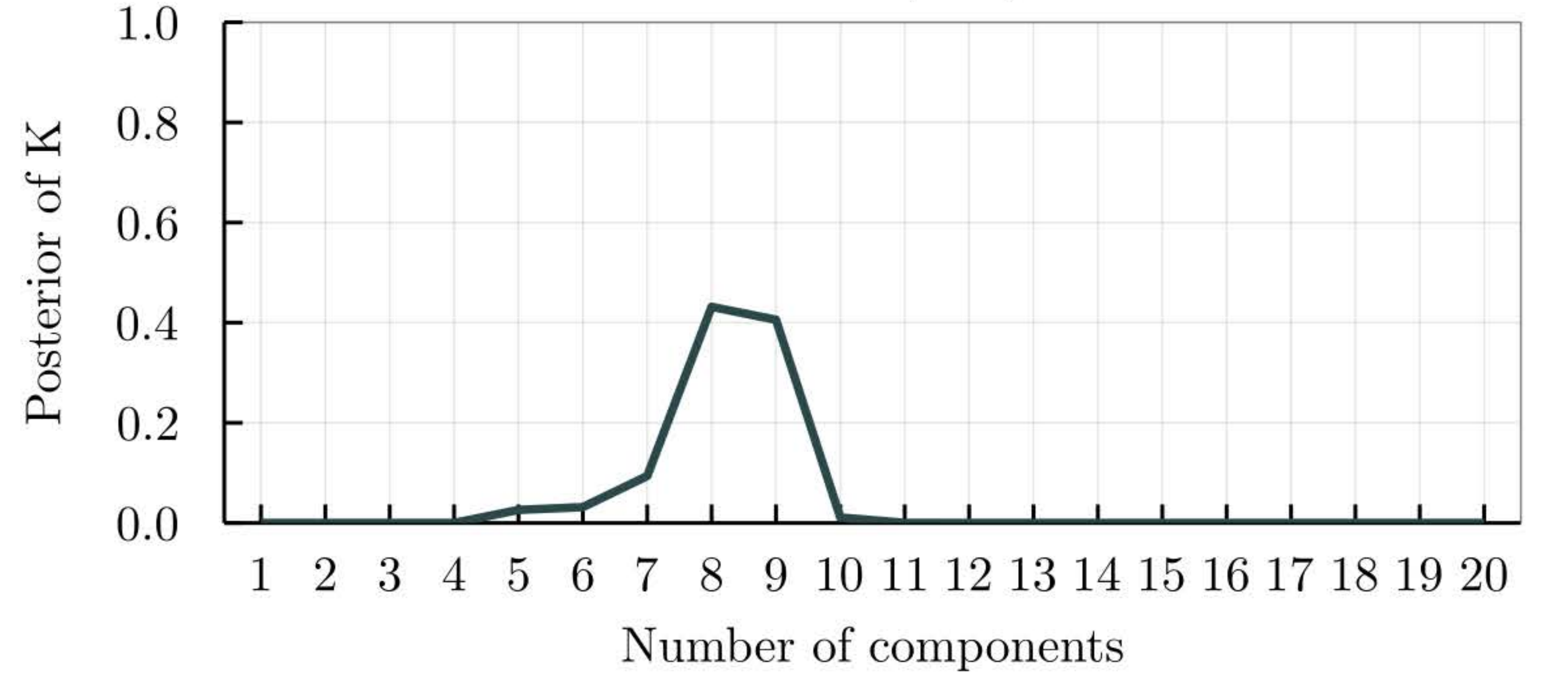}
        \subcaption{MFM (full covariance model).}
    \end{subfigure}
    \begin{subfigure}{0.49\linewidth}
        \includegraphics[width=1\linewidth]{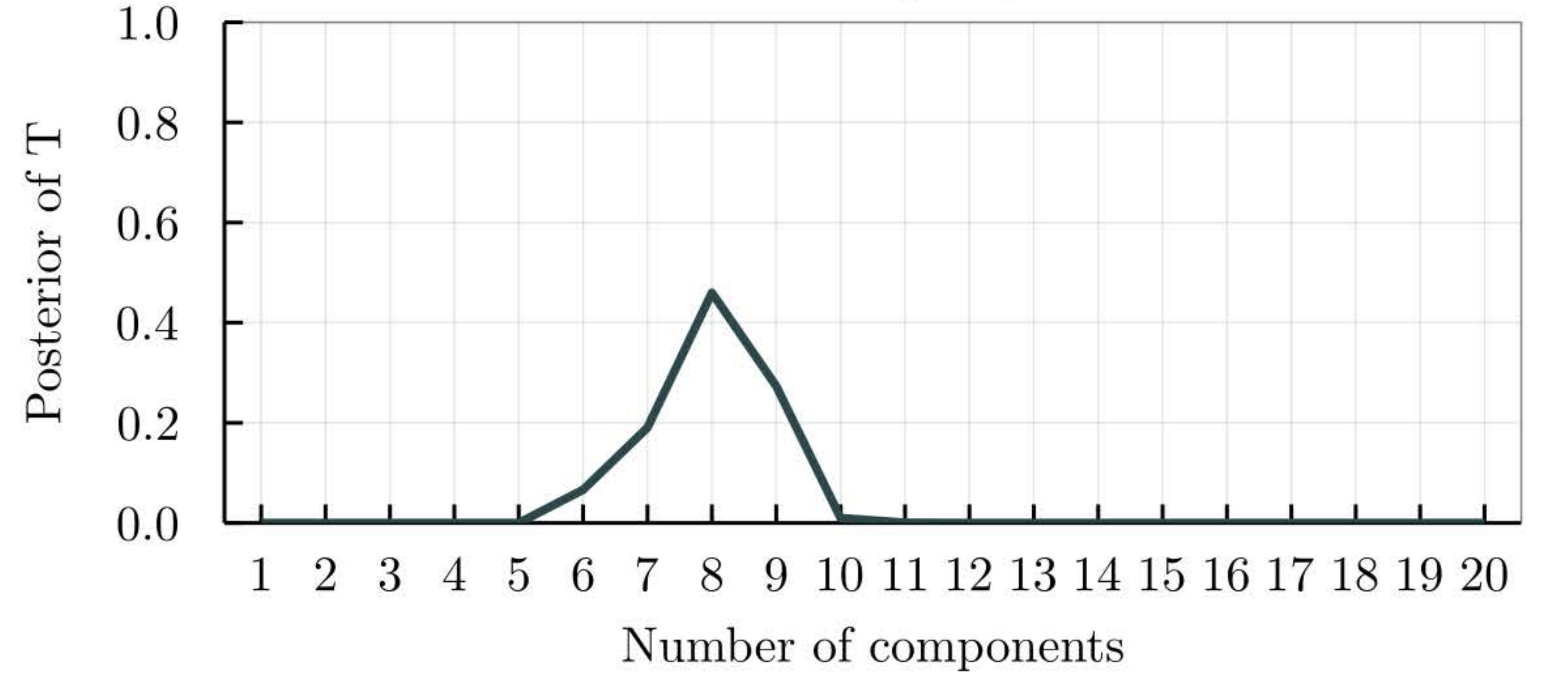}
        \subcaption{DPM (full covariance model).
        }
    \end{subfigure}
    
    \bigskip
    
    \begin{subfigure}{0.49\linewidth}
        \includegraphics[width=1\linewidth]{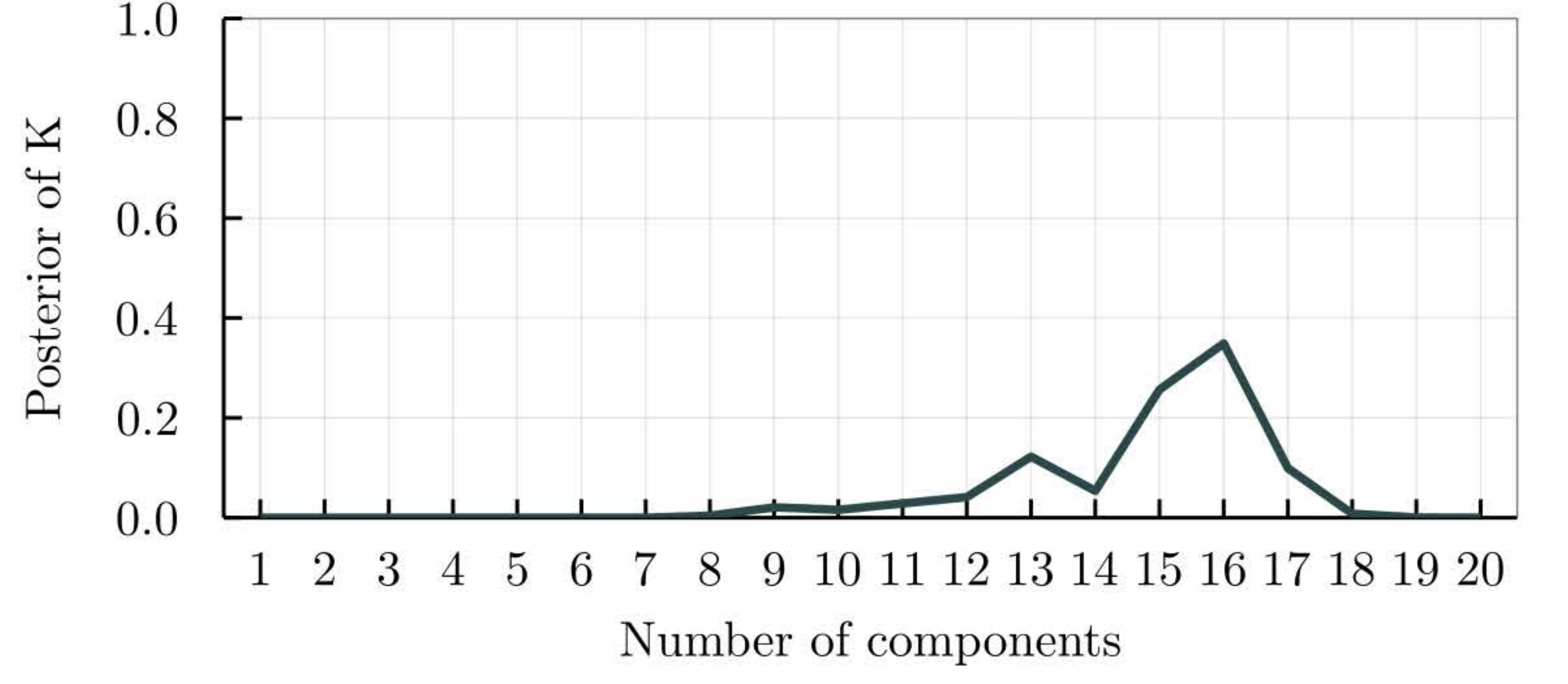}
        \subcaption{MFM (diagonal covariance model).}
    \end{subfigure}
    \begin{subfigure}{0.49\linewidth}
        \includegraphics[width=1\linewidth]{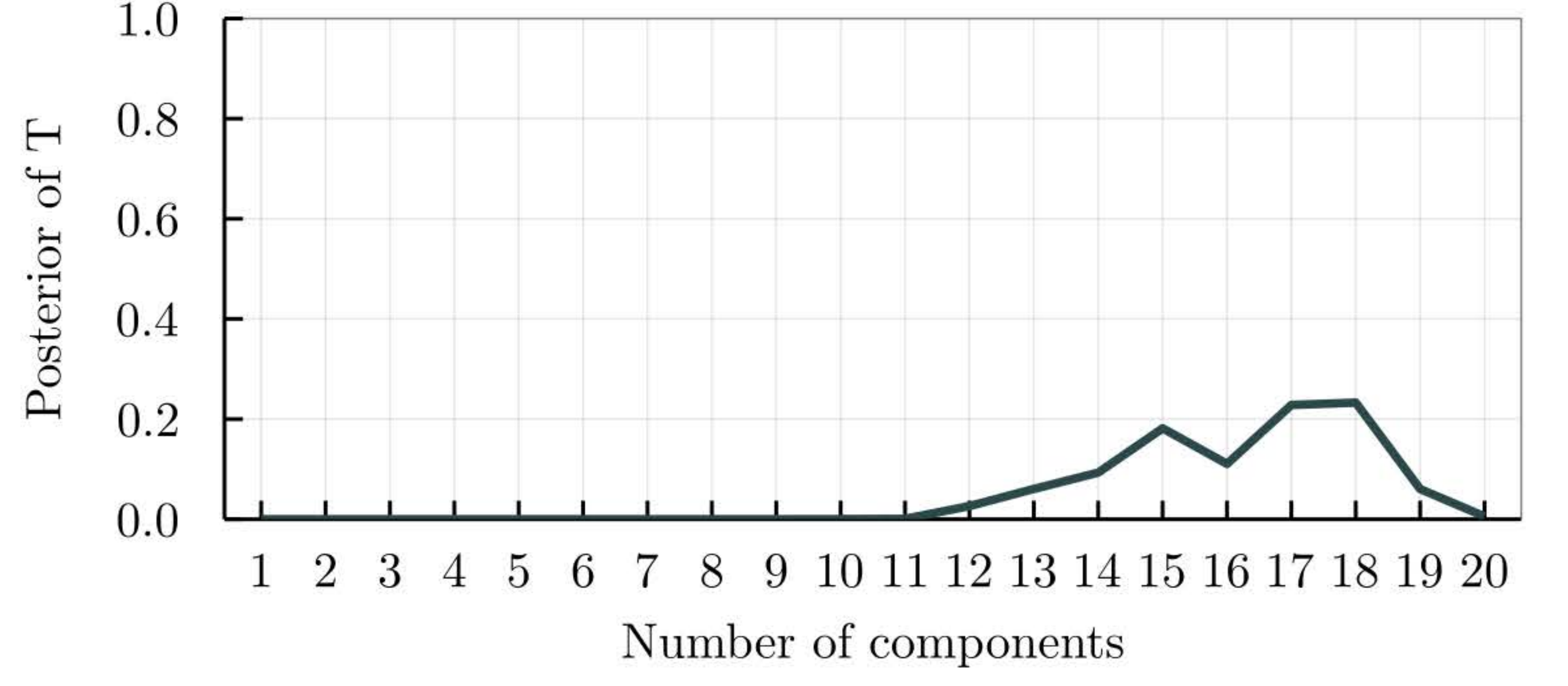}
        \subcaption{DPM (diagonal covariance model).
        }
    \end{subfigure}    \caption{Posterior numbers of components for the mixture of finite mixtures (MFM) and Dirichlet process mixture (DPM) with full or diagonal covariance on mouse single-cell RNA-sequence data. There are seven true clusters according to \cite{zeisel15}.}
    \label{fig:mouse_posts}
\end{figure}

Figure~\ref{fig:mouse_posts} gives the posterior distributions of the number of components for models \eqref{def:mvn} and \eqref{def:mvnaac}.
The MFM and DPM models give similar posteriors for the full covariance model, with a mode at eight components, suggesting that the DPM overestimation of the number of components tends to be moderate for sample sizes in the order of $10^3$ and confirming our results in Section~\ref{subsec:res_dpm}.
The MFM posterior gives slightly more credibility to higher numbers of components than does that of the DPM.  The diagonal covariance specification leads to a significant overestimation of the number of components, confirming the theoretical results from Section~\ref{subsec:miss} and our simulation results in Section~\ref{subsec:res_miss}.
Furthermore, the DPM model contributes to the overestimation of the number of components, with a mode at 18 components compared to 16 for the MFM model.
Both model types show bimodal posterior distributions with a second local mode and have similar shapes, though the DPM posterior is shifted towards higher numbers of components.

% \begin{figure}
%     \centering
%     \includegraphics[width=1\linewidth]{figures/gene/posts.png}
%     \caption{}
%     \label{fig:mouse_posts}
% \end{figure}

Table~\ref{tab:gene_nclust} gives the number of clusters of summarisation methods. We do not consider methods based on partitioning around medoids due to computational limitations. Overall we confirm our simulation results, with overestimation of the number of clusters for Binder and PEAR losses with hierarchical clustering, especially with average linkage.  Medvedovic clustering and variation of information correct the overestimation of the Dirichlet process mixture model with diagonal covariance, giving numbers of clusters similar to the MFM model. However, as in Section~\ref{subsec:res_miss}, summarisation methods do not correct the overestimation due to misspecification.  The adjusted Rand indices show that the full covariance model is more accurate than the misspecified diagonal covariance model.  We cannot expect perfect recovery of the true seven cell types, even with the full covariance model, because we use a very general model without consideration of the data type and without further prior assumptions.

\begin{table}
    \centering
    \begin{tabular}{ V{1.5} c V{1.5} c c c c V{1.5} c c c c V{1.5} }
    \multicolumn{1}{c}{} & \multicolumn{4}{c}{\textbf{Number of clusters}} &  \multicolumn{4}{c}{\textbf{Adjusted Rand index}} \\
    \clineB{2-9}{1.5}
     \multicolumn{1}{cV{1.5}}{} & \multicolumn{2}{c}{Full} &  \multicolumn{2}{cV{1.5}}{Diagonal} 
     &  \multicolumn{2}{c}{Full} &  \multicolumn{2}{cV{1.5}}{Diagonal}\\
    \multicolumn{1}{cV{1.5}}{} & \multicolumn{1}{c}{MFM} & DPM & MFM & DPM & MFM & DPM & MFM & DPM \\
    \hlineB{1.5}
    
    Medvedovic & 8 & 8 & 15 & 16 & 0.38 & 0.35 & 0.34 & 0.35 \\
    \rowcolor{gray!15}
    Binder+Average & 30 & 36 & 59 & 55 & 0.38 & 0.38 & 0.33 & 0.33 \\
 
    Binder+Complete & 11 & 12 & 19 & 17 & 0.38 & 0.38 & 0.34 & 0.35 \\
    \rowcolor{gray!15}
    Binder+Samples & 9 & 9 & 15 & 18 & 0.37 & 0.38 & 0.34 & 0.34 \\
  
    PEAR+Average & 25 & 29 & 33 & 32 & 0.38 & 0.38 & 0.33 & 0.33 \\
    \rowcolor{gray!15}
    PEAR+Complete & 11 & 12 & 18 & 17 & 0.38 & 0.38 & 0.34 & 0.35 \\

    PEAR+Samples & 9 & 9 & 15 &  18 & 0.37 & 0.38 & 0.34 &  0.34 \\
    \rowcolor{gray!15}
    VI-LB+Average & 9 & 8 & 14 & 15 & 0.38 & 0.38 & 0.33 & 0.35 \\

    VI-LB+Complete & 9 & 9 & 15 & 15 & 0.38 & 0.38 & 0.34 & 0.35 \\
    \rowcolor{gray!15}
    VI-LB+Samples & 9 & 8 & 15 & 19 & 0.37 & 0.37 & 0.34 & 0.33 \\
    \hlineB{1.5}
    Posterior mode & 8 & 8 & 16 & 18 & \multicolumn{4}{c}{} \\
    
    \cellcolor{gray!15} Truth  & \cellcolor{gray!15}7 & \cellcolor{gray!15}7 & \cellcolor{gray!15}7 & \cellcolor{gray!15}7 & \multicolumn{4}{c}{} \\
    \clineB{1-5}{1.5}
    \end{tabular}
    
    \caption{Number of clusters and adjusted Rand indices of summary clusterings for the MFM and DPM models with full or diagonal covariance on mouse single-cell RNA-sequence data. The adjusted Rand index compares the summary clustering with the true clustering with seven clusters given by \cite{zeisel15}.}
    \label{tab:gene_nclust}
\end{table}

\begin{figure}
    \centering
    \hspace*{-20mm}
    \begin{subfigure}{0.49\linewidth}
        \includegraphics[width=1\linewidth]{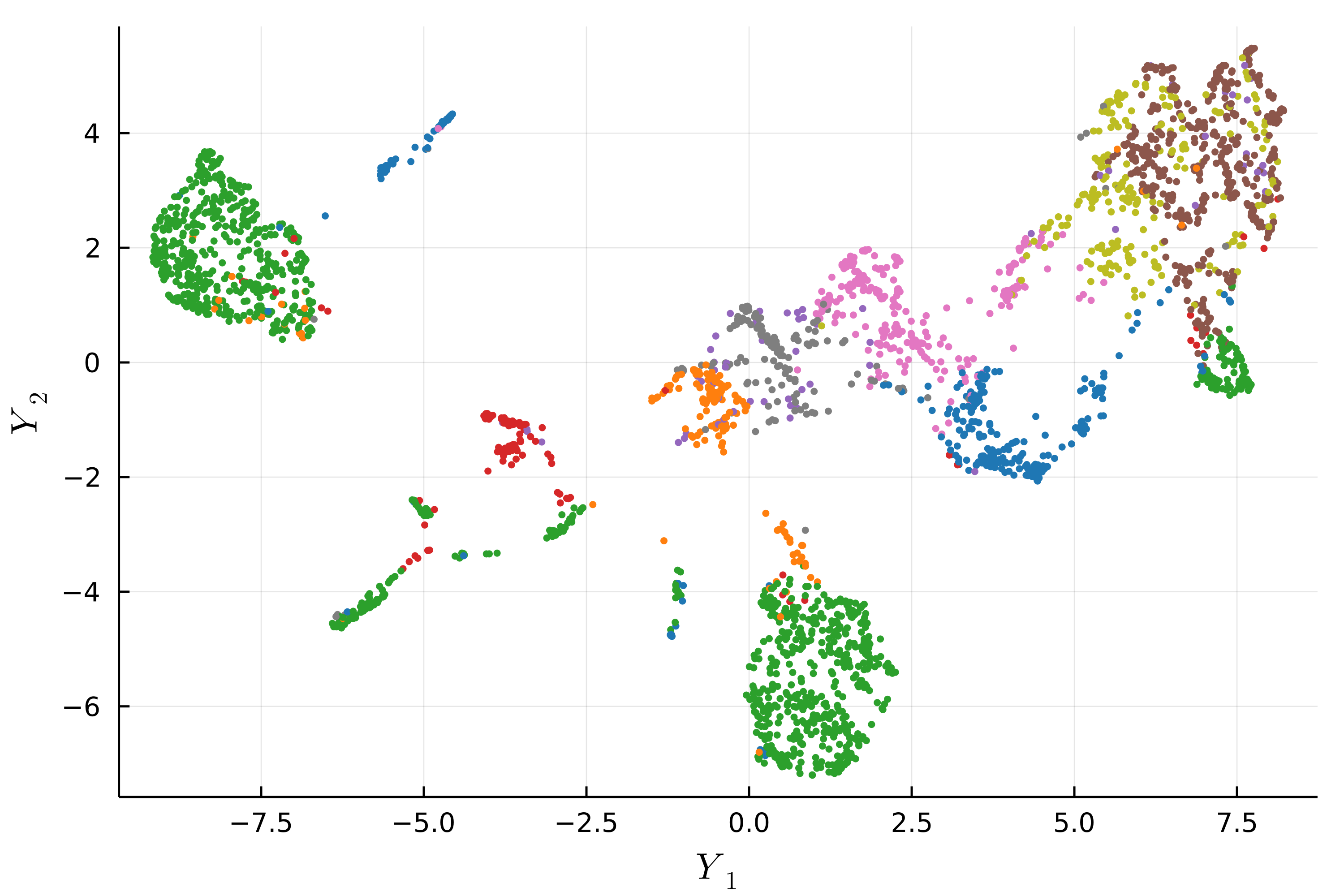}
        \subcaption{Full covariance model.\label{subfig:umap_mvn}}
    \end{subfigure}
    \hspace{-2mm}
    \begin{subfigure}{0.49\linewidth}
        \includegraphics[width=1\linewidth]{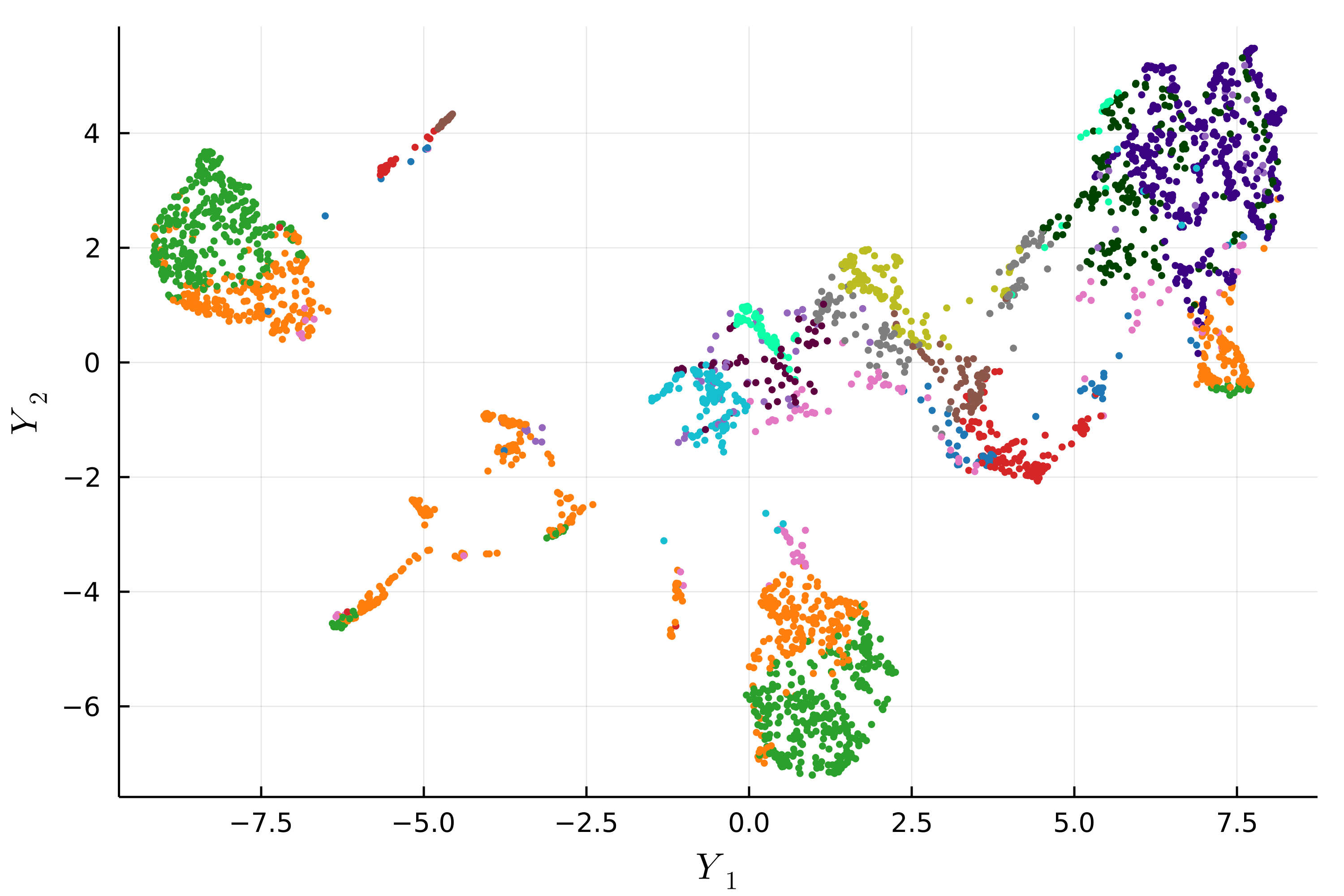}
        \subcaption{Diagonal covariance model.\label{subfig:umap_mvnaac}}
    \end{subfigure}
    \begin{subfigure}{0.49\linewidth}
        \includegraphics[width=1\linewidth]{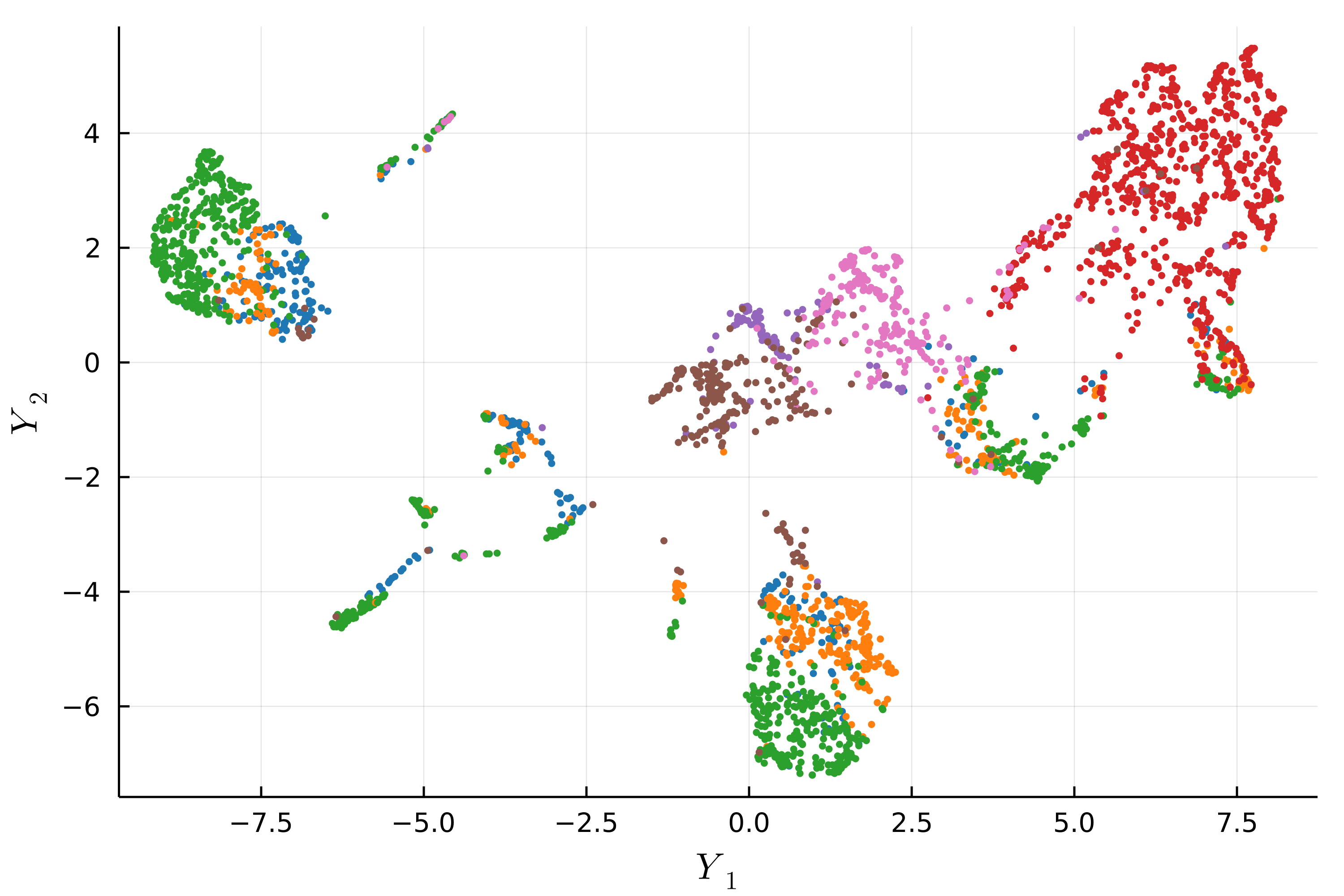}
        \subcaption{True clusters.\label{subfig:umap_true}}
    \end{subfigure}
    \caption{Summary clusterings of variation of information average linkage hierarchical clustering for the MFM model with a full (a) or diagonal (b) covariance specification on the mouse single-cell RNA-sequence data. We visualise the 10-dimensional data with the Uniform Manifold Approximation and Projection (UMAP) on two dimensions, denoted by $( Y_1, Y_2)$. The true clusters (c) are given by \cite{zeisel15}.}
    \label{fig:gene_umap}
\end{figure}

\begin{figure}
    \centering
    \includegraphics[width=0.65\linewidth]{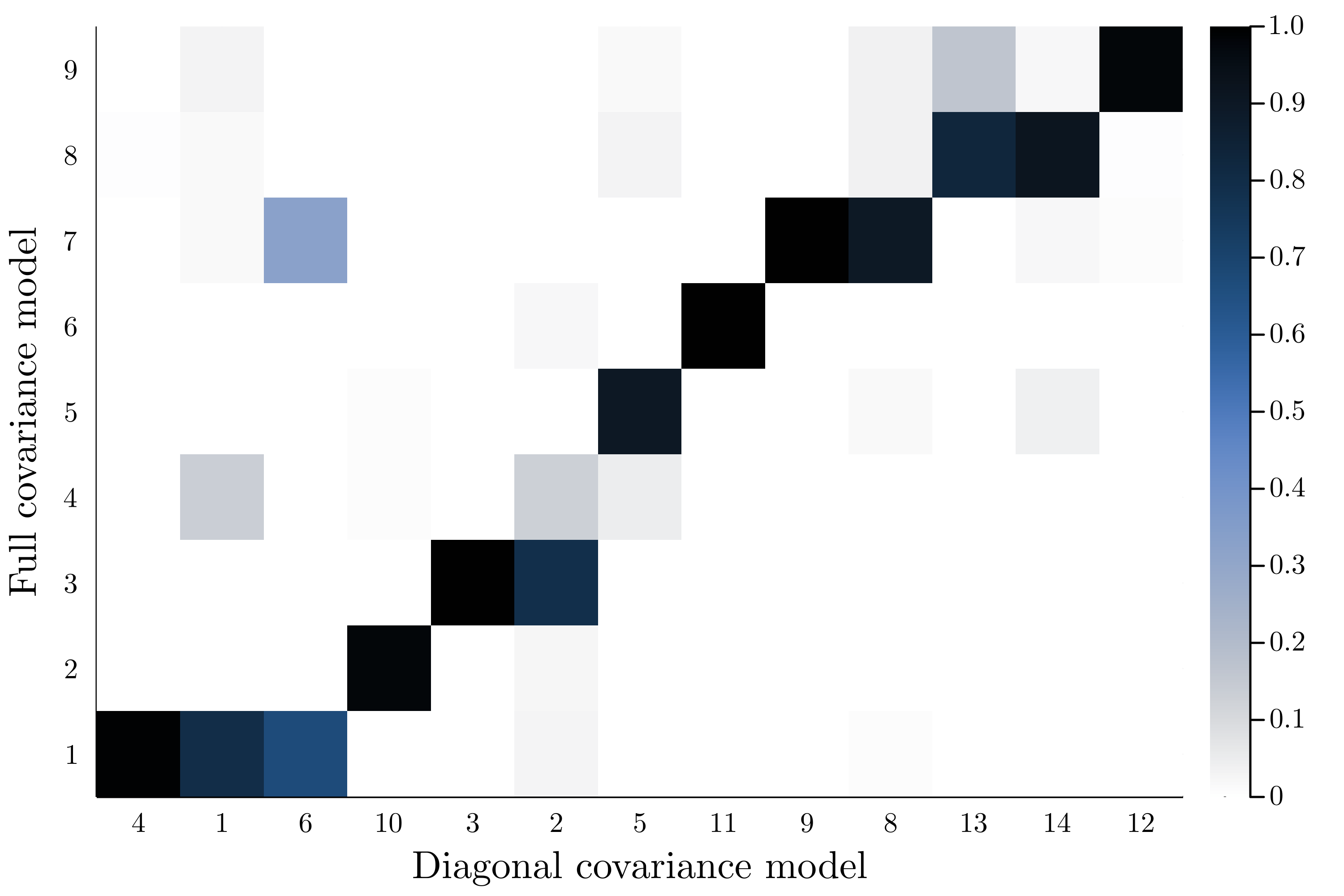}
    \caption{Normalised contingency counts between summary clusterings of the MFM model with a full and diagonal covariance on the mouse single-cell RNA-sequence data. We reorder clusters of the diagonal covariance by their largest count and normalise columns for readability, so that entries correspond to proportions relative to the diagonal covariance cluster sizes. We use variation of information average linkage hierarchical clustering for both models.}
    \label{fig:gene_cont}
\end{figure}

We use the Uniform Manifold Approximation and Projection (UMAP) \citep{mcinnes20} to visualise our clusterings. Figure~\ref{fig:gene_umap} illustrates summary clusterings for the MFM model on a two-dimensional projection of our data, which we compare to the true clustering according to \cite{zeisel15}.
The misspecified model seems to create subclusters of the full covariance, i.e., the clusters of the diagonal covariance model seem to split up the larger clusters of the full covariance model. For synthetic data, we showed in Section~\ref{subsec:res_miss} that the diagonal covariance model splits up clusters with correlated data. We confirm our conjecture in Figure~\ref{fig:gene_cont} by visualising the contingency counts for summary clusterings of the MFM model. By reordering the clusters of the diagonal covariance model, we roughly split up clusters of the full covariance model (rows) into clusters of the diagonal covariance (columns). For example, cluster 1 of the full diagonal covariance model can be split into clusters 4, 1 and 6 of the diagonal covariance model. 

In conclusion, our analysis of the mouse single-cell RNA-sequence data confirms some of our simulated results. Misspecification of the component distributions seems to have a bigger effect on the overestimation of the number of components than does the Dirichlet process prior. Summarisation methods can correct the overestimation of the DPM model compared to the MFM model but they do not correct for misspecification of the component distributions. However, misspecification of Gaussian component distributions seems to split components of their well-specified counterpart, giving reasonably interpretable summary clusterings.

\section{Discussion}
\label{sec:conclusion}

Our simulation results confirm theoretical inconsistencies for the number of components of Bayesian mixture models from recent literature for finite samples. MCMC summarisation methods may correct estimation errors, but the summarisation method matters, as some popular methods can lead to severe overestimation. Our gene expression analysis supports these findings and shows the importance of appropriate model specification.

Dirichlet process mixture models overestimate the number of components asymptotically in the sample size \citep{miller14}, but  our results show limited overestimation for practical sample sizes. Moreover, some summarisation methods consistently correct overestimation and give the correct number of clusters. In our simulations, for example, variation of information \citep{wade18} consistently led to accurate summary clusterings. We believe that summarisation methods can find the true number of clusters because additional clusters introduced by the DPM model tend to be small and diffused across MCMC samples.
We therefore expect that past studies using this model are mostly unaffected by these theoretical drawbacks, especially if they use summary clusterings for final interpretations.
Nevertheless, we recommend the mixture of finite mixtures model in general, due to its theoretical consistency and equally efficient sampling methods.

Bayesian mixture models with misspecified component distributions overestimate the number of components asymptotically in the sample size \citep{cai21}, and our results show that this overestimation can be large for Gaussian mixture models, and cannot be corrected by summary clusterings.
Nonetheless, we conjecture that misspecified mixture models tend to only split true clusters, leading to moderate errors in interpretation.  However, our results are limited to misspecification by constraining component distributions to be less expressive, rather than specifying entirely different families of distributions.

Some summary clustering methods lead to overestimation of the number of clusters, even for well-specified models and simple examples, where MCMC samples concentrate around the true number of clusters. We have shown that summary clusterings obtained with Binder's loss \citep{binder78} or the posterior expected average adjusted Rand index \citep{fritsch09} overestimate the true number of clusters on synthetic and real data, when using common optimisation methods. These results agree with those from simulations in \cite{rastelli18} and \cite{wade18}, where they also observed overestimation for Binder's loss (using equal penalties of misclassification) and correct estimation by the lower bound of the variation of information loss. Moreover, we have shown that the choice of optimisation methods for summary clustering can greatly influence the number of clusters. Here we have compared some of the most popular methods, but acknowledge that there are many such methods \citep[e.g., more recently,][]{rastelli18,dahl21} and hence our analyses are not exhaustive. By exploring a bigger part of the space of all possible partitions, these more recent methods are able to further minimise the loss function, but interestingly this seems to have a negative impact on the estimation of the number of clusters. 

Our results only consider Gaussian mixture models, so we could extend our work to other continuous models, such as Laplace location mixtures or regression models, or to discrete mixture models, such as multinomial mixture models. Such extensions, however, would require other Gibbs updates for split-merge sampling. An alternative would be to generate data from a different family of distributions and keep using Gaussian mixture models.

Our results show that the summary number of clusters can greatly depend on the summarisation method, so theoretical work might help explain under what conditions different methods perform best.

Finally, when misspecification is chosen over efficiency, we could investigate guided summarisation methods that account for misspecification by merging additional clusters. Introducing intuitive adjustable settings to summarisation methods would accommodate further assumptions more generally, and allow the user to choose  appropriate representative clusterings  according to the purpose of subsequent analysis.

% Acknowledgements should go at the end, before appendices and references

\acks{PDWK and JvdMM acknowledge MRC grant MC\_UU\_00002\_13.  This work was supported by the Swiss National Science Foundation and the National Institute for Health Research [Cambridge Biomedical Research Centre at the Cambridge University Hospitals NHS Foundation Trust]. The views expressed are those of the authors and not necessarily those of the NHS, the NIHR or the Department of Health and Social Care.  }


\newpage

\appendix
\section*{Appendix A.}

\begin{figure}[!h!]
    \centering
    \includegraphics[width=1.0\linewidth]{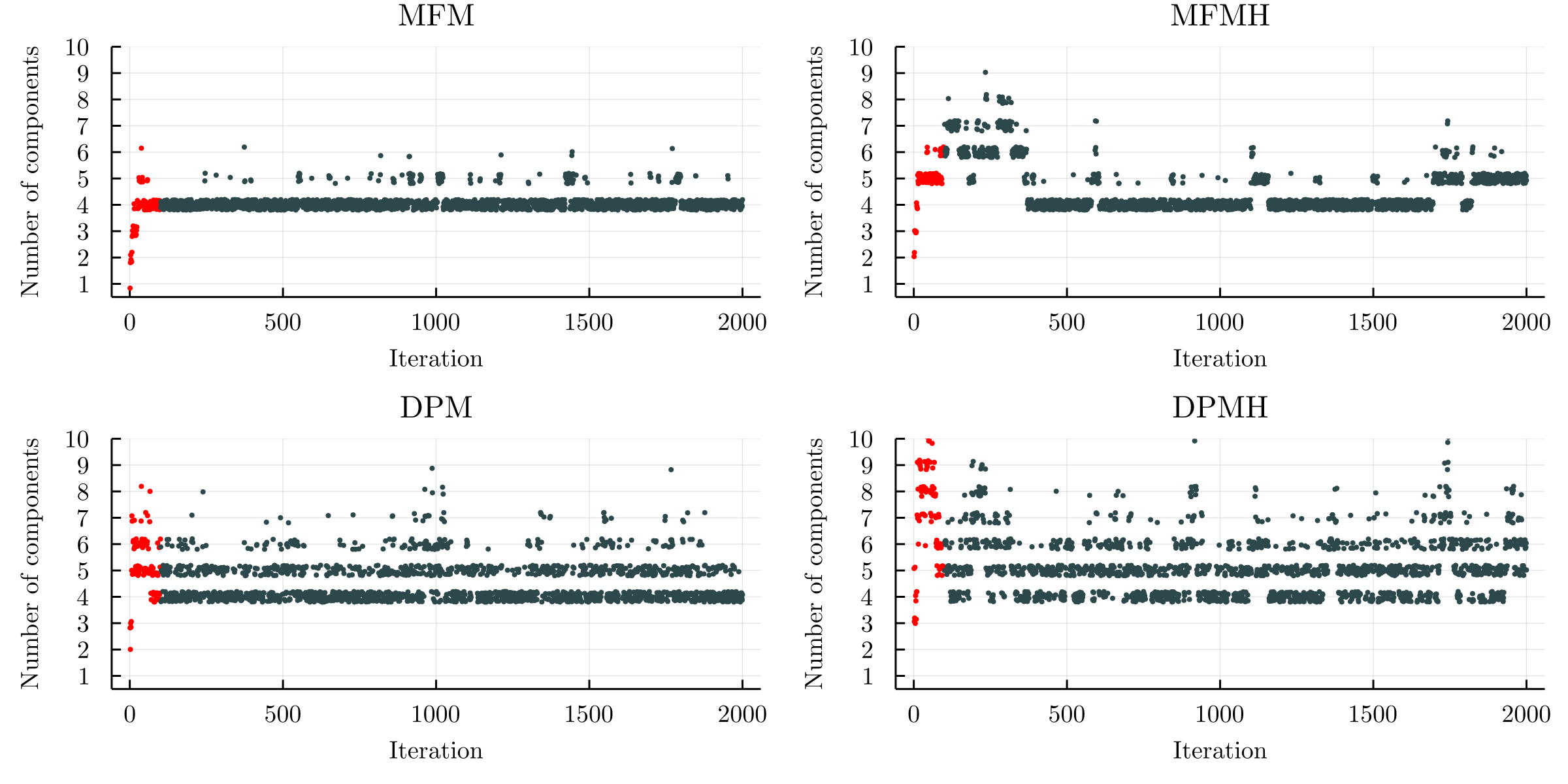}
    \caption{Number of components of MCMC sample allocation for a mixture of finite mixtures with (MFMH) and without (MFM) a hyperprior and for a Dirichlet process mixture with (DPMH) and without (DPM) a hyperprior, jittered for visibility, on synthetic data with $N = 500$. The 100 samples in red are burn-in and not included in posterior analysis.}
    \label{fig:example_trace}
\end{figure}

%\begin{figure}[!h!]
%    \centering
%    \includegraphics[width=1.0\linewidth]{figures/example/example_p%ost.png}
    %\caption{Posterior number of components for a mixture of finite %mixtures with (MFMH) and without (MFM) a hyperprior and for a %Dirichlet process mixture with (DPMH) and without (DPM) a %hyperprior on synthetic data with $N = 500$ and four generating %components.}
    %\label{fig:example_post}
%\end{figure}
\label{app:theorem}

% \begin{figure}[!h!]
%     \centering
%     \includegraphics[width=0.6\linewidth]{figures/example/example_psm.png}
%     \caption{Example of a posterior similarity matrix for the mixture of finite mixtures (MFM) model without a hyperprior on synthetic data with $N=500$ and four components. The posterior similarity between two observations is the posterior co-clustering probability estimated from MCMC samples. Observations are reordered with hierarchical clustering to improve interpretability.}
%     \label{fig:example_psm}
% \end{figure}

\begin{figure}
    \centering
    \includegraphics[width=1\linewidth]{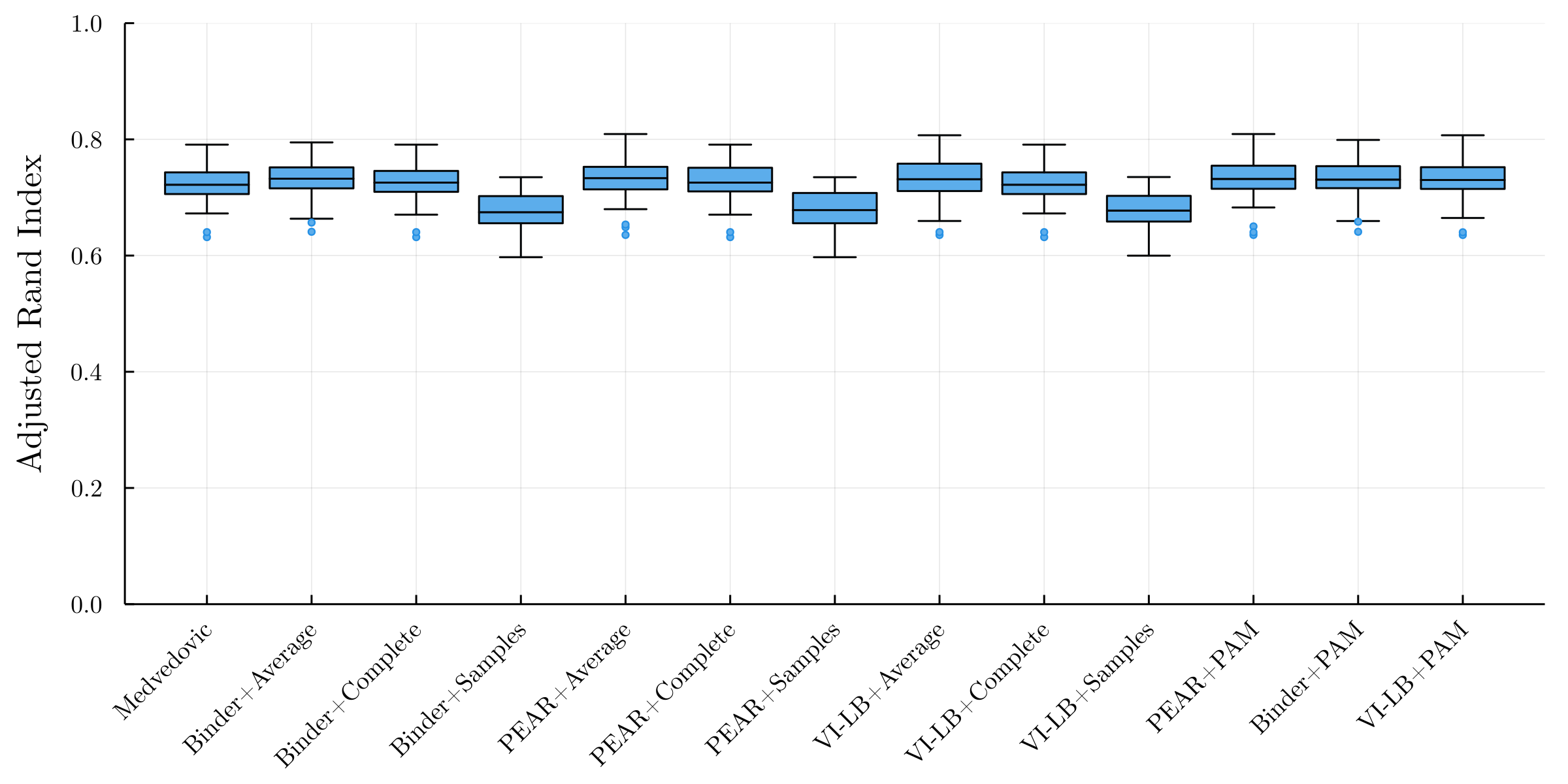}
    \caption{Adjusted Rand index of summary clusterings with the true generating cluster allocation for the MFM model without a hyperprior on $50$ independent synthetic datasets with $N = 500$. Summary names are composed of the loss function and the optimisation method, except for Medvedovic clustering.}
    \label{fig:example_ari}
\end{figure}

\begin{figure}
    \centering
    \includegraphics[width=0.8\linewidth]{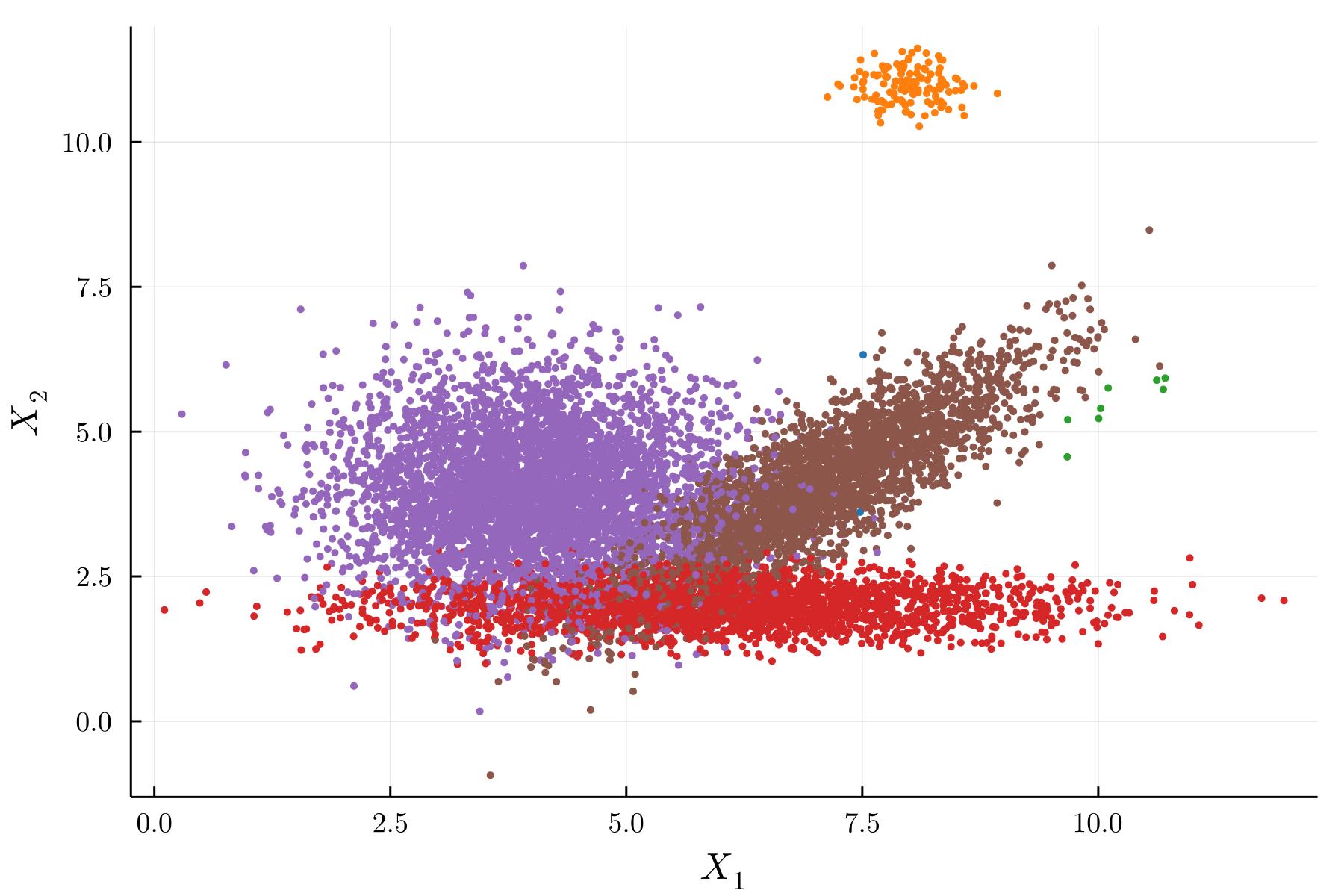}
    \caption{Example of a MCMC allocation sample for the Dirichlet process mixture model with a hyperprior on synthetic data with $N= 10^4$. Observations are coloured by their cluster allocation. One can discern six clusters, including a green one on the right and a blue one roughly aligned on $X_1 = 7.5$.}
    \label{fig:dpm_samp}
\end{figure}

\begin{figure}
    \centering
    \begin{subfigure}{0.49\linewidth}
        \includegraphics[width=1\linewidth]{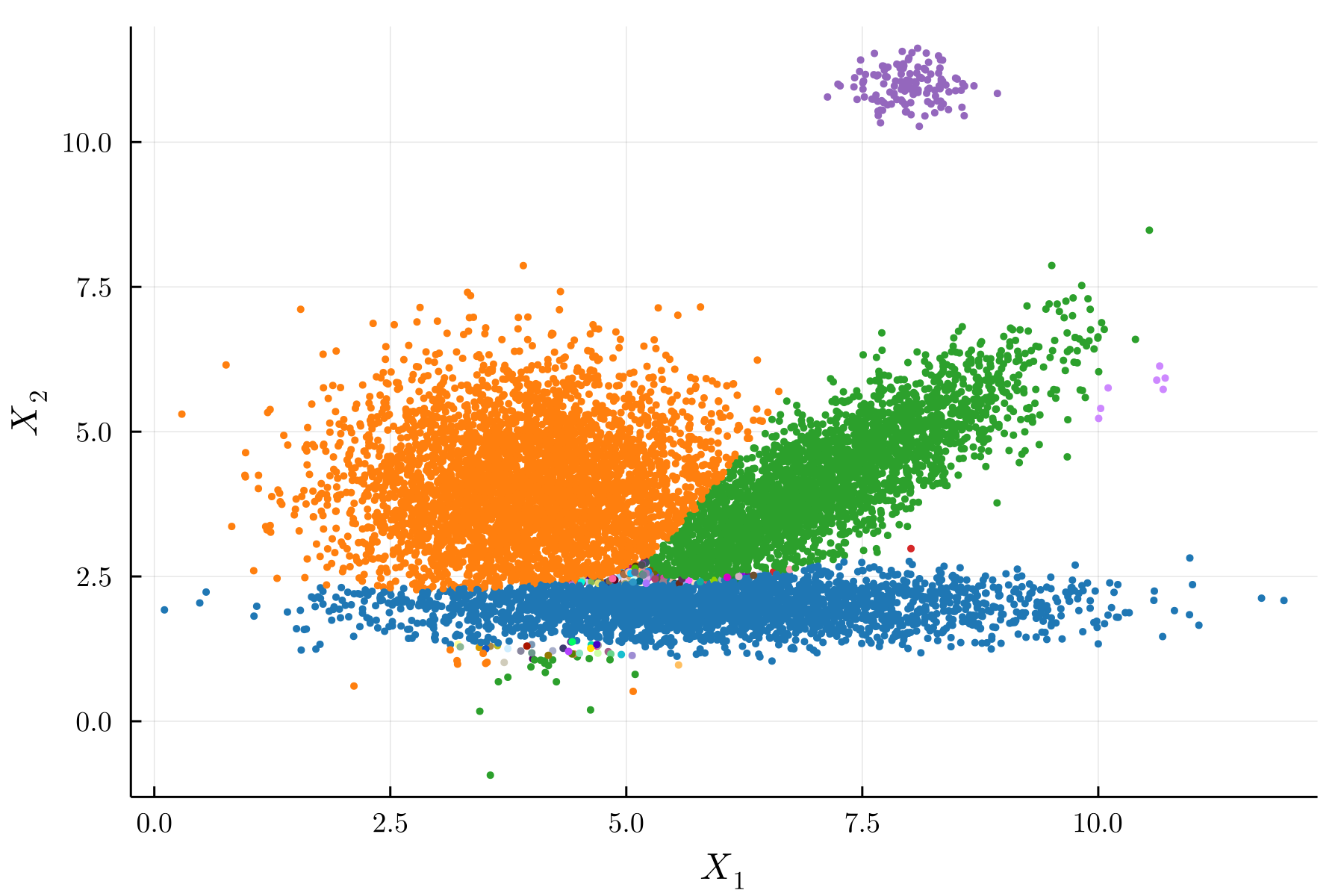}
        \subcaption{}
    \end{subfigure}
    \hfill
    \begin{subfigure}{0.49\linewidth}
        \includegraphics[width=1\linewidth]{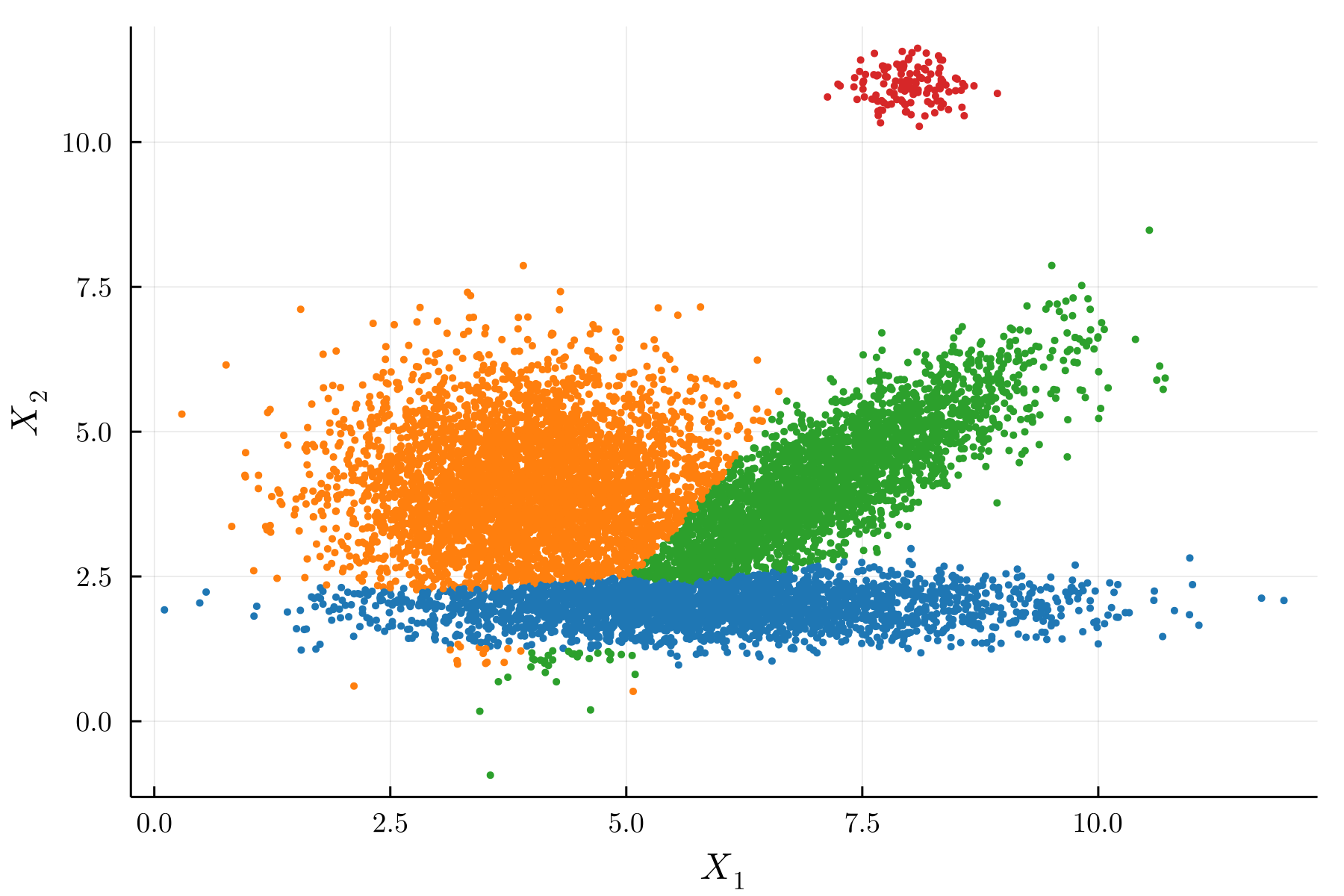}
        \subcaption{}
    \end{subfigure}
    \caption{Summary clusterings of PEAR loss (a) and variation of information (b) both optimised with average linkage hierarchical clustering on MCMC samples of the DPM model with a hyperprior on synthetic data with $N=10^4$.}
    \label{fig:dpm_clust}
\end{figure}

% Note: in this sample, the section number is hard-coded in. Following
% proper LaTeX conventions, it should properly be coded as a reference:

%In this appendix we prove the following theorem from
%Section~\ref{sec:textree-generalization}:

\end{document}